\documentclass[aapm,graphicx]{revtex4-1}

\draft 

\usepackage[cmex10]{amsmath}
\usepackage{tabularx}
\usepackage{multirow}
\usepackage{booktabs}
\usepackage{color}
\usepackage{array}

   \usepackage[pdftex]{graphicx}
   \DeclareGraphicsExtensions{.pdf,.jpeg,.png,.eps}
\usepackage{epstopdf}

\usepackage[normalem]{ulem}

\begin{document}


\title{A Model-Based Scatter Artifacts Correction for Cone Beam CT} 



\author{Wei~Zhao$^{1}$,~Don Vernekohl$^{2}$,~Jun~Zhu$^{1}$,~Luyao~Wang$^{1}$,~Lei Xing$^{2}$}%
\email[]{lei@stanford.edu}
\affiliation{$^{1}$ Department of Biomedical Engineering, Huazhong University of Science and Technology, Hubei, China.}
\affiliation{$^{2}$ Stanford University, Department of Radiation Oncology, Stanford, CA 94305 USA.}



\begin{abstract}

\textbf{Purpose:} Due to the increased axial coverage of multi-slice computed tomography (CT) and the introduction of flat detectors, the size of X-ray illumination fields has grown dramatically, causing an increase in scatter radiation. For CT imaging, scatter is a significant issue that introduces shading artifact, streaks, as well as reduced contrast and Hounsfield Units (HU) accuracy. The purpose of this work is to provide a fast and accurate scatter artifacts correction algorithm for cone beam CT (CBCT) imaging.\\
\textbf{Methods:} The method starts with an estimation of coarse scatter profiles for a set of CBCT data in either image domain or projection domain. A denoising algorithm designed specifically for Poisson signals is then applied to derive the final scatter distribution. Qualitative and quantitative evaluations using thorax and abdomen phantoms with Monte Carlo (MC) simulations, experimental Catphan phantom data, and \emph{in vivo} human data acquired for a clinical image guided radiation therapy were performed. Scatter correction in both projection domain and image domain were conducted and the influences of segmentation method, mismatched attenuation coefficients and spectrum model as well as parameter selection were also investigated.\\
\textbf{Results:} Results show that the proposed algorithm can significantly reduce scatter artifacts and recover the correct HU in either projection domain or image domain. For the MC thorax phantom study, four components segmentation yield the best results, while the results of three components segmentation are still acceptable. The parameters (iteration number $K$ and weight $\beta$) affect the accuracy of the scatter correction and the results get improved as $K$ and $\beta$ increase. It was found that variations in attenuation coefficient accuracies only slightly impact the performance of the proposed processing. For the Catphan phantom data, the mean value over all pixels in the residual image is reduced from -21.8 HU to -0.2 HU and 0.7 HU for projection domain and image domain, respectively. The contrast of the \emph{in vivo} human images are greatly improved after correction.\\
\textbf{Conclusions:} The software-based technique has a number of advantages, such as high computational efficiency and accuracy, and the capability of performing scatter correction without modifying the clinical workflow (i.e., no extra scan/measurement data are needed) or modifying the imaging hardware.  When implemented practically, this should improve the accuracy of CBCT image quantitation and significantly impact CBCT-based interventional procedures and adaptive radiation therapy.
\end{abstract}

\pacs{}

\maketitle 

\section{Introduction}

Scatter contamination remains to be one of the most important problems in cone beam CT (CBCT) imaging. Usually, artifacts are present when the model used in the reconstruction algorithm is not consistent with the projection data acquisition model. Despite of extensive efforts from the imaging science community, existing reconstruction algorithms in clinically used CBCT systems do not model the scatter radiation adequately, leading to severe scatter artifacts and hindering the maximal utilization of the technology. Indeed, scatter artifacts often manifest themselves as shading or streaks between high contrast objects, reduced contrast resolution, and inaccurate Hounsfield Units (HUs). Scatter correction has been extensively studied in the past decades but a clinically reasonable solution remains illusive. Current scatter correction methods can be briefly classified into five approaches: physical scatter rejection, analytical modeling, Monte Carlo (MC) simulation, primary modulation, and scatter measurements.

Physical scatter rejection techniques employ an air gap, an anti-scatter grid, or bow-tie filter in the data acquisition systems~\cite{Mail2008,Ruhrnschopf2011,Schafer2012,Sisniega2013}. They usually yield insufficient correction and additional scatter correction is recommended. Additionally, with the introduction of an anti-scatter grid, soft-tissue contrast-to-noise ratio (CNR) may be affected due to the inevitable concomitant rejection of primary events. As thus an undesired increase in dose would be necessary to achieve prior image quality~\cite{siewerdsen2004a,Schafer2012}.

In analytical modeling methods, a scatter potential which is usually a function of primary signals convolved with a scatter kernel is employed to estimate the scatter radiation distribution in the measured raw data~\cite{Naimuddin1987,Boone1988,Seibert1988,Ohnesorge1999,rinkel2007,Li2008,Maltz2008,Star-Lack2009,Sun2010,Meyer2010a,Star-Lack2013,sun2014,kim2015,bhatia2016}. These methods preserve the field-of-view (FOV), and require no extra hardware and additional scan. Computationally they are efficient with a predefined kernel. The achievable accuracy of the methods depends, however, heavily on the reliability of the model used for scatter artifacts correction.
 MC simulation based methods, in which the scatter kernel~\cite{Swindell1996,Hansen1997,Spies2001,Maltz2008,Reitz2009,Wiegert2010, bootsma2015} or the scatter radiation distribution  ~\cite{Kyriakou2006,Jarry2006,Bertram2008,Mainegra-Hing2008,Colijn2004,Zbijewski2006,Mainegra-Hing2010,Poludniowski2009,xu2014,thing2014,bootsma2015} is computed directly by following through the trajectories of all involved photons, provide a more robust modeling of the photon transport process. The technique is, however,  computationally expensive and its potential for routine clinical use remains questionable. For this reason, Monte Carlo simulation is often used in combination with other approaches~\cite{meyer2010,Baer2012,Star-Lack2013}.

Both spatial and temporal primary modulation methods have been studied for scatter correction. The former~\cite{Maltz2006,Zhu2006,gao2010a,gao2010b,ritschl2015} assumes that scattered photons are predominantly low frequency in their spatial distribution. With the presence of a primary beam modulator, the primary signals are separated from the scatter signals in the Fourier domain. On the other hand, the scatter profile in the latter approach is assumed to be unchanged by the temporal modulation of the primary modulator~\cite{Schorner2012}. The approach relies on the use of a demodulation method to estimate the primary signal. A drawback of these methods is that they require some extra hardware support or mechanical modification of the scanners.

Measurement-based scatter correction estimates scatter signals from blocked areas in partially blocked X-ray beam profiles~\cite{Ning2004,Siewerdsen2006,Liu2006,Zhu2009,Jin2010,Yan2010,Niu2011a,Lee2012,min2015}. In this approach, the scatter photon distribution is extracted by measuring the signals in periodically shadowed regions of a beam modifying device placed between the X-ray source and patient, under an underlying assumption that the scatter signal is predominately low frequency in space~\cite{Zhu2009}. A pre-scan with the beam modifier in place and with less view angles than that of a norm scan is usually required~\cite{Ning2004,Zhu2009,Jin2010}. To avoid double scans, scatter fluence estimation from pixel values near the edge of the detector behind the collimators or in the "wing" region was attempted~\cite{Siewerdsen2006,Meng2013,lu2015}. An advanced method using compressed sensing optimization algorithm is also proposed to estimate the scatter profile with the signals from the edges of the field of view~\cite{Meng2013}. Furthermore, the use of moving blockers to avoid double scanning was investigated for improved CBCT image quality~\cite{Zhu2005,Wang2010,Ouyang2013}. In addition to the above methods, post-corrections using basis images and a level set method were proposed to mitigate scatter induced cupping artifacts~\cite{Meyer2010a,xie2015}.

In this work, a novel scatter correction method for cone beam CT (CBCT) imaging is investigated. The essence of the approach is a coarse-to-fine estimation of the scatter signals by effectively utilizing the useful features of the system at various stages of the calculation. Briefly, the calculation starts with a rough estimation of the scatter profiles for a given set of data in either image domain or projection domain. A denoising algorithm designed specifically for Poisson signals is then applied to refine the scatter profiles to derive the final scatter distribution. Extensive validation of the proposed approach is carried out using MC simulations, phantom measurements, and human data. Our results demonstrate that the proposed approach is robust and works well in various testing situations. The software-based technique has a number of advantages, such as high computational efficiency and accuracy, and the capability of performing scatter correction without modifying the clinical workflow (i.e., no extra scan/measurement data are needed) or modifying the imaging hardware.  When implemented practically, this should improve the accuracy of CBCT image quantitation and significantly impact CBCT-based interventional procedures and adaptive radiation therapy.

\section{Methods and materials}
Measured projection data are comprised of primary and scatter signals. A major task to remove the adverse effects of scatters is to find the scattered radiation distribution. The true signal is obtained by subtracting the scatter distribution from the measured raw data, i.e.
\begin{equation}
I_{p}(\alpha,\vec{x}) = I(\alpha,\vec{x})-I_{s}(\alpha,\vec{x}),
\end{equation}
where $I_{p}$ is the corrected projection data, namely, the estimated primary projection data, $I$ is the raw data and $I_s$ is the scatter signal. Indices $\alpha $ and $\vec{x}$ stand for projection view angle and detector channel number, respectively. For simplicity, we will drop the indices in the following descriptions.

\begin{figure}[t]
     \centering
     \includegraphics[width=3.5in]{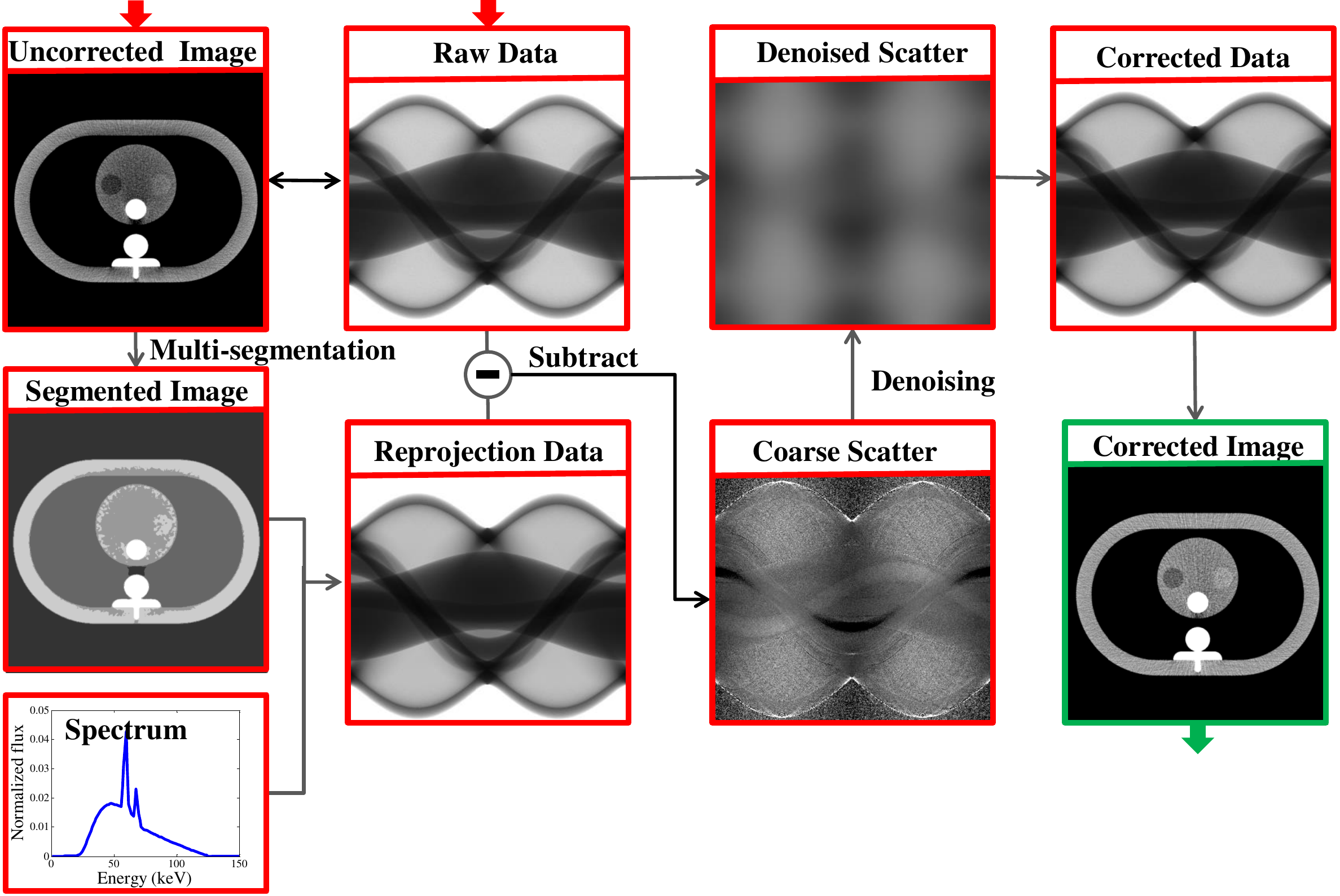}
     \caption{Flowchart of the proposed scatter correction method. The method can start with either the raw projection data or a scatter contaminated volume as input.}
     \label{fig:f1}
 \end{figure}

The proposed method, as shown in Fig.~\ref{fig:f1}, starts with generating a coarse scatter by polychromatic reprojection (Section A). To obtain the coarse scatter, scatter contaminated CBCT images reconstructed using the raw projection data, were segmented and a polychromatic reprojection of the segmented images is performed with consideration of the predetermined X-ray spectrum (Section B) and known data acquisition geometry. The reprojected data is then subtracted from the measured raw projection data at each given view angle to generate a coarse estimate of the scatter profile for the subsequent denoising algorithm (Section C).

\subsection{Coarse scatter estimation}

The uncorrected CT image is first segmented using the OTSU method~\cite{otsu1975} which exhaustively search the optimum thresholds separating the image into different classes so that their intra-class variance is minimal. Based on the segmented image volume, the primary signal can be modeled as follows:
\begin{equation}\label{equ:reporjection}
\hat{I}_{p}=N\int_{0}^{E_{max}}\mathrm{d}E\,\Omega(E) \, \eta(E)\,\mathrm{exp}\left[-\int_{0}^{l}\mu(E,s)\mathrm{d}s\right],
\end{equation}
where $N$ is the total number of photons, $\Omega(E)$ the polychromatic X-ray spectrum, and $\eta(E)$ the energy-dependent efficiency of the detector.
$E_{max}$ is the maximum photon energy of the spectrum. $\mu(E,s)$ is the energy-dependent linear attenuation coefficient and $l$ is the propagation path length for each ray, and is calculated using a GPU-based ray-tracing algorithm~\cite{pratx2011,jia2014}. Using these notations, the flood field $I_{0}$ is written as:
\begin{equation}\label{equ:floodfield}
I_{0}=N\int_{0}^{E_{max}} \mathrm{d}E\,\Omega(E)\,\eta(E).
\end{equation}
After replacing $N$ with $I_{0}$ and substituting it into (\ref{equ:reporjection}), the result describes the estimated primary signal with flood field $I_{0}$:
\begin{equation}\label{equ:primary}
\hat{I}_{p}=\frac{I_{0}\int_{0}^{E_{max}}\mathrm{d}E\,\Omega(E)\,\eta(E)\,\mathrm{exp}\left[-\int_{0}^{l}\mu(E,s)\mathrm{d}s\right]}{\int_{0}^{E_{max}}\mathrm{d}E\,\Omega(E)\,\eta(E)}.
\end{equation}
Subtracting the above estimated primary signal from the measured total projection data $I$,  the coarse scatter signal $\hat{I}_s$ becomes:
\begin{equation}\label{equ:coarseScatter}
\hat{I}_{s}=I-\frac{I_{0}\int_{0}^{E_{max}}\mathrm{d}E\,\Omega(E)\,\eta(E)\,\mathrm{exp}\left[-\int_{0}^{l}\mu(E,s)\mathrm{d}s\right]}{\int_{0}^{E_{max}}\mathrm{d}E\,\Omega(E)\,\eta(E)}.
\end{equation}
Note that the segmented image volume is generated from a scatter contaminated reconstruction, which is in turn used to estimate the primary signals in our method.

\subsection{Polychromatic spectrum estimation}

To accurately calculate the primary signal $\hat{I}_{p}$, the energy spectrum $\Omega(E)$ used in the polychromatic reprojection should be modeled precisely. In this study, an indirect transmission measurement-based spectrum estimation method was employed to estimate an effective spectrum which can model the polychromatic attenuation process of the projection data~\cite{zhao2015}. The technique is briefly summarized below.

The method starts with the reconstruction of a volume using the raw projection data. The first step of the spectrum estimation is to segment the uncorrected images into different components using OTSU method~\cite{otsu1975}. By calculating the propagation path length (PPL) for each of the segmented components for each detector pixel, we generate a set of polychromatic reprojection data using (\ref{equ:reporjection}) with the PPLs and an estimated polychromatic spectrum. The estimated spectrum is then iteratively updated to minimize the difference of the measured and reprojected data. To further improve the robustness of the iterative spectrum estimation procedure, the estimated spectrum is expressed as a weighted summation of a set of model spectra $\Omega_{i}(E)$ which are obtained using either Monte Carlo simulation or analytical spectrum generators~\cite{siewerdsen2004b,poludniowski2009a} with different filtration, i.e. the estimated spectrum $\Omega(E)$ can be expressed as follows,
\begin{equation}\label{equ:spek}
\Omega(E)=\sum_{i=1}^{M}c_{i}\Omega_{i}(E),
\end{equation}
with $M$ the number of the model spectra, $c_{i}$ the unknown weights. Based on the model spectra expression, the spectrum estimation problem is formulated as the following iterative optimization problem,
\begin{equation}\label{equ:opt-constraint}
\mathbf{c}=\mathrm{argmin}_{\mathbf{c}}\;\|p_{\mathrm{m}}-\hat{p}(\mathbf{c})\|_{2}^{2}, ~~\mathrm{s.t.}~\sum_{1}^{M}c_{i}=1,~\mathrm{and}~c_{i}>0.
\end{equation}
Here $p_{\mathrm{m}}$ is the measured projection and $\hat{p}$ is the polychromatic reprojection and it is a function of the unknown weights $\mathbf{c}$, with both taken the logarithmic operation. Within the study only projections $p_{\mathrm{m}}$ from experimental setups with negligible scatter contaminations are used for the spectrum estimation itself, as good spectrum estimations are required. The normalization constraint condition $\sum_{1}^{M}c_{i}=1$ and the non-negative constraint condition $c_{i}>0$ are used to normalize the estimated spectrum to unit area and to keep the solution of (\ref{equ:opt-constraint}) physically meaningful, respectively. Note that the detector response which can be regarded as the multiplication of photon energy and the absorption efficiency for an energy integrating detector~\cite{zhao2015}, is included during the spectrum estimation.

\subsection{Denoising the coarse scatter}

Since the coarse scatter estimate $\hat{I}_s$ is dependent on the segmentation procedure, it may yield inaccurate results, especially for low contrast objects that have similar attenuation properties as the background material and for the edges of two neighboring materials. To compensate for the inaccuracy caused by segmentation, instead of regularizing the coarse scatter using a convolution-based scatter model~\cite{zhao2014,zhao2015scatter}, in this study, we directly denoise the coarse scatter (the residual of the raw projection data and the polychromatic reprojection data) using a statistical-based denoising algorithm.

It is well known that scatter signal is predominantly low frequency in both spatial and temporal domains. By assuming that the scatter photons $I_s$ arrive at a specific pixel under \textit{Poisson} distributed statistics, it is possible to recover the noise-free scatter projection $I_s$ from the coarse scatter $\hat{I}_s$ by using denoising techniques specifically designed for signals that consider the \textit{Poisson} distributed origin. Such a possible algorithm to yield a smooth scatter distribution from $\hat{I}_s$ was presented in~\cite{jia2012} and is used in this study. The denoising algorithm is aimed to solve the following optimization problem,
\begin{equation}\label{equ:energyfunction}
I_s=\mathrm{argmin}_{I_s}\;\int \mathrm{d}\vec{x}(I_s-\hat{I}_s \mathrm{log}I_s)+\frac{\beta}{2}\int d\vec{x}|\nabla I_s|^{2}.
\end{equation}
Note that $I_s$ and $\hat{I}_s$ are detector pixel dependent and the integration will run over all of the detector pixels. The first term of (\ref{equ:energyfunction}) is a data-fidelity term that considers the \textit{Poisson} statistics and keeps $I_s$ close to the data $\hat{I}_s$, while the second one is a regularization term to keep the solution $I_s$ smooth, namely, being dominated with low frequency content. $\beta$ is a constant to determine the relative weight of the two terms. This objective function is convex and can be solved using a variational approach~\cite{le2007,jia2012}.

The final numerical calculation expression suitable for implementation using an iterative algorithm (the successive over-relaxation algorithm is employed in this study) can be formulated as follows~\cite{jia2012}:
\begin{equation}\label{equ:implementation}
\begin{split}
I_s^{(k+1)}(i,j)=&(1-\omega)I_s^{(k)}(i,j)+\\
& \frac{\omega}{4}[\sum I_s^{(k)}(i,j)-\frac{1}{\beta}(1-\frac{\hat{I}_s(i,j)}{I_s^{(k)}(i,j)})].
\end{split}
\end{equation}

Here $(i,j)$ is the detector pixel location index. $\sum I_s(i,j)$ stands for $I_s(i+1,j)+I_s(i,j+1)+I_s(i-1,j)+I_s(i,j-1)$ and $k$ is the iteration number. $\omega$ is an empirical value and $\omega=0.8$ is used during the processing. 

\begin{figure}[t]
     \centering
     \includegraphics[width=3.5in]{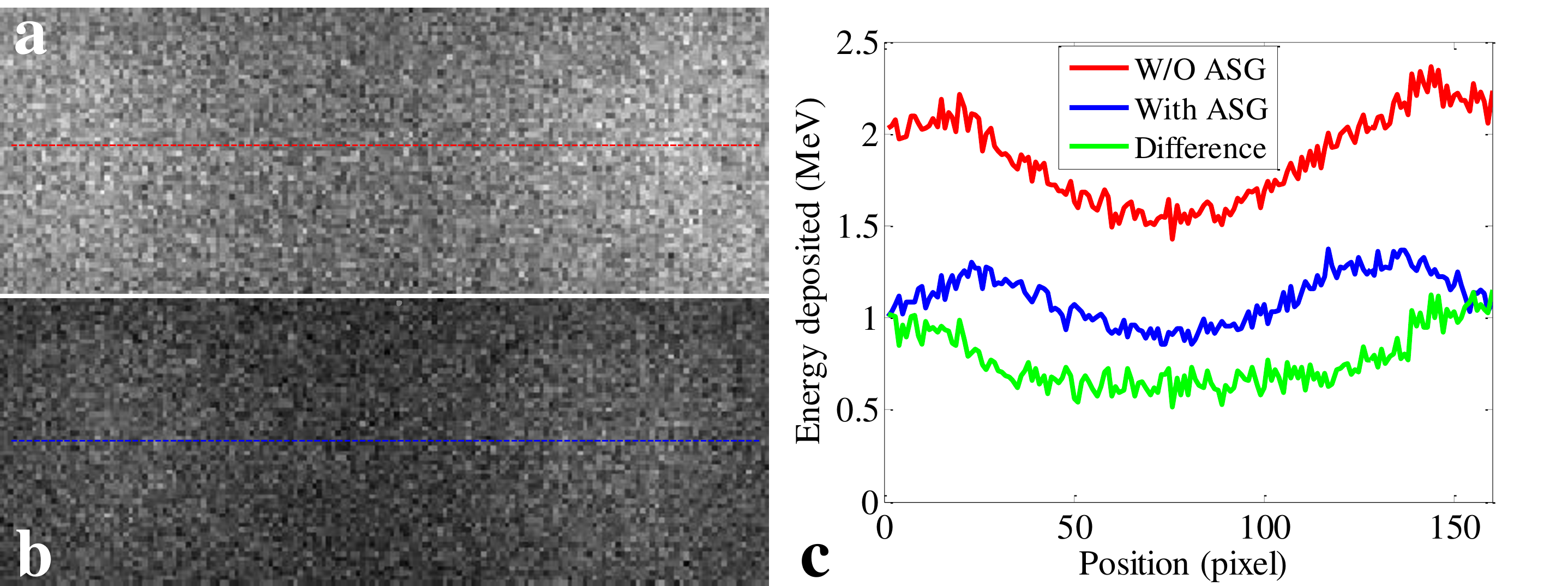}
     \caption{Profiles of the MC simulated scatter radiation of an abdomen phantom with and without anti-scatter grid (ASG) using MC simulation. (a) without ASG, (b) with a 1D ASG (grid density-80lp/cm, grid ratio-6:1). (c) Scatter radiation line profiles without ASG (red), with ASG (blue) and their difference (green). }
     \label{fig:scatterASG}
 \end{figure}

The denoising above relies on the low frequency feature of the scatter radiation. However, an anti-scatter grid (ASG) may be employed in a realistic application and it is unclear whether the scatter is still dominated by low frequency content in this case. Fig.~\ref{fig:scatterASG} shows profiles of the scatter distributions of an abdomen phantom with and without ASG. It is visible that the magnitude of the scatter distribution decreases significantly when the ASG is used. However, the global profile of the scatter signal with ASG does not change significantly, namely, the scatter radiation with ASG is still dominantly in the low frequency domain as that without ASG.

In addition, some of the commercial CBCT scanners have incorporated built-in scatter correction procedures, such as the kernel-based analytical modeling. In this case, the low frequency feature of the scatter radiation is also employed and the scatter signal calculated using the built-in methods are generally smooth signals. Thus, it is assumed that the residual scatter signal, which is the difference between the true scatter signal and the scatter signal pre-calculated using the built-in methods, is also a smooth signal and can be applied with the denoising procedure.

\subsection{Implementation of the method in image domain}

The proposed method can also be applied on scatter artifacts contaminated images or a volume as input. In some clinical scanners, no access to the appropriate raw projection data format is provided and therefore scatter correction can only be accomplished in image domain. Based on the fact that tomographic reconstruction is a linear process, i.e. the order of summation and backprojection operations are interchangeable, a scatter projection error $\Delta p_s$, which can be added linearly in the logarithmic raw-data domain, is first calculated. Let $p_c$ be the scatter corrected projection value after taking logarithm, its value is given by
\begin{equation}
p_c=\mathrm{ln}\frac{I_0}{I-I_{s}}.
\end{equation}
Since we have $p_c=p+\Delta p_s$ and the logarithmic raw projection data $p=\mathrm{ln}(I_0/I)$, $\Delta p_s$ can be expressed as
\begin{equation}\label{equ:deltaScatter}
\Delta p_s=\mathrm{ln}\frac{1}{1-\widetilde{I}_s e^{p}},
\end{equation}
where $\widetilde{I}_s = I_{s}/I_0$. In (\ref{equ:deltaScatter}), $p$ can be obtained by a forward projection of the uncorrected images, thus in order to get $\Delta p_s$, we need to calculate $\widetilde{I}_s$, i.e. the pre-denoised $\widetilde{\hat{I}}_s=\hat{I}_{s}/I_0$. Meanwhile, for each of the pre-logarithmic projection data $I$, we have $I=I_0 e^{-p}$. After substituting this expression into (\ref{equ:coarseScatter}), the pre-denoised $\widetilde{\hat{I}}_s$ can be calculated as follows,
\begin{equation}\label{equ:preDenoisedScatter}
\widetilde{\hat{I}}_s=e^{-p}-\frac{\int_{0}^{E_{max}}\mathrm{d}E\,\Omega(E)\,\eta(E)\,\mathrm{exp}\left[-\int_{0}^{l}\mu(E,s)\mathrm{d}s\right]}{\int_{0}^{E_{max}}\mathrm{d}E\,\Omega(E)\,\eta(E)}.
\end{equation}
From this, we can generate the denoised scatter estimate $\widetilde{I}_s$ and now all of the variables in (\ref{equ:deltaScatter}) are known or can be pre-calculated. We then reconstruct $\Delta p_s$ to obtain the scatter error images or volume $\Delta f_s$. By simply adding the $\Delta f_s$ to the uncorrected raw images or volume $f_u$, we obtain the scatter corrected image $f_c$,
\begin{equation}\label{equ:imageDomainSAC}
f_c=f_u+\Delta f_s.
\end{equation}
Note that these images are functions of spatial variable and (\ref{equ:imageDomainSAC}) is operated in a pixel-wise fashion. It should be noted that based on the assumption of the subsection above, for the image domain implementation, the input uncorrected image may be an image that is preprocessed using the built-in methods and residual shading artifacts are still presented.

\subsection{Monte Carlo simulations}

To validate the proposed algorithm, an anthropomorphic 
thorax phantom and an abdomen phantom were used to generate MC simulation data with the Geant4-based MC simulation package GATE~\cite{Jan2011}. To quantitatively and qualitatively investigate the effect of scatter artifacts reduction, a water insert including three small low contrast compartments (adipose, breast and liver) is placed in the central area of the two phantoms. A bone insert is also included in the phantom. In the GATE simulation of X-ray CT scanning, in order to store the targeted events information, two types of sensitive detectors (the crystalSD and phantomSD) were defined. Physical interactions including photo-electric, ionization, \textit{Compton}-, \textit{Rayleigh}- and multiple scattering within the sensitive detectors are recorded. The crystalSD is attached to the CT detector to score the photons that arrived at the detector, while the phantomSD is attached to the CT phantom to retrieve information about the \textit{Compton-} and the \textit{Rayleigh}-scattered events within the phantom. For any volume attached to the sensitive detector, information (such as energy deposition, geometrical information, position and time, types of interaction, etc.) are stored. The object oriented data analysis framework ROOT is used to extract and analyze  the recorded photon events. 

For the analysis, primary projection data, scatter only projection data, and primary plus scatter projection data were extracted independently. The primary photon events were defined as photons which are scored by the crystalSD and did not undergo interactions in the phantom. The scattered photon events were defined as photons which are scored by the crystalSD and have at least one \textit{Compton-} or \textit{Rayleigh-} scatter interaction in the phantom. Since energy-integrating detectors are commonly used in clinical applications, total energy deposited in a specific time interval and at a specific detector pixel is considered as the projection data of the specific pixel in the specific view angle. One of the advantages of using MC simulations is that they can differentiate primary photons and scatter photons. The primary plus scatter projection data, referred to as the total projection data, correspond to the realistic projection data acquired using a CBCT scanner and are corrected using the proposed algorithm. Thus the estimated scatter signal and the corrected image can be compared directly to the true scatter signal and the primary image (reconstructed using primary projection data and served as the ground truth), respectively.

For computational purpose, a parallel geometry and a plane X-ray source (2D $320\times120$~mm$^{2}$ rectangle source) were used in the MC simulations. The distance from the source to the center of rotation is 750 mm and the distance from the detector to the center of rotation is 450 mm. A circular scan was simulated where a total of 360 projections per rotation are acquired over an angular range of $360^{\circ}$. The detector element size is $1\times36$~mm$^{2}$ (width $\times$ height) and the detector consists of 320 detector columns. Since the correction is performed in projection domain, the reprojection geometry should be the same with the MC geometry, namely, a parallel reprojection geometry. The material of the detector elements is CsI and its thickness is 1 mm. For clinical CBCT scanners, the thickness of the CsI crystals in the flat-detectors is usually 600~$\mu m$. We slightly increased the crystal thickness to 1~mm here to improve the absorption efficiency. The polychromatic X-ray source spectrum is 125 kVp and it is generated using the Spektr software~\cite{siewerdsen2004b} with 5 mm aluminum filtration. For each of the simulations, a total number of $3\times10^{10}$ photons were emitted and the whole simulation time took about one week on a computer node machine which equipped 32 cores of Opteron 6134.

Since scatter radiation effects the spectrum estimation, the 5 mm aluminum filtered 125 kV polychromatic spectrum was first recovered using a water cylinder phantom in 2D setup. In this case, the effect of scatter radiation is negligible. During spectrum estimation, ten Spektr~\cite{siewerdsen2004b} model spectra were used and the hardest spectrum was employed as initial guess for the optimization problem Eq.~(\ref{equ:opt-constraint}).

\subsubsection{Calculation in projection and image domain}

We first investigate the scatter correction for the anthropomorphic thorax phantom and the abdomen phantom in projection and image domain. Primary projections and primary plus scatter (total) projections were used to reconstruct the primary images and the total images, respectively, using a FBP algorithm with the band-limited Ramp filter (i.e. Ram-Lak filter) whose
cut-off frequency is set to the Nyquist frequency. 
During the correction, the denoising was performed with $\beta$=100 and $K$=500. A quantitative analysis on the corrected images was applied to test the accuracy of the HUs for different regions of interest (ROIs) in the water insert by comparing them to the values of the primary image.

\subsubsection{Segmentation}

The polychromatic reprojection is performed using the uncorrected CBCT images with all relevant structures (such as adipose, soft tissue and bone) segmented. However, due to the limited contrast resolution, structures with similar attenuation coefficient may not be differentiated from each other during the segmentation, which may affect the final results. To evaluate the influence of segmentation, different segmentation methods (two-, three- and four-component segmentation) were performed for the same uncorrected CBCT images. A polychromatic reprojection was then performed for each of the segmented images. For all of the scenarios, $\beta$ and $K$ are set to 10000 and 1700, respectively. The resulting scatter corrections were then evaluated in accordance to the accuracy of HUs for the different segmentation methods. Furthermore, we compared the results to a perfect segmentation where each component was well identified.

\subsubsection{Parameters selection}

In order to optimize the denoising parameters $\beta$ and iteration number $K$ in (\ref{equ:implementation}), scatter corrections in both projection and image domain using different $\beta$ and $K$ values were performed for the anthropomorphic thorax phantom. The HU accuracy for different ROIs were used again to quantitatively and qualitatively characterize the quality of the corrections.

\subsubsection{Robustness evaluation}

As mentioned above, an attenuation coefficient needs to be assigned to each of the segmented components. The attenuation values are usually obtained from the NIST database and then interpolated for all energies between 0 and 150 keV with the spline method. These values are regarded as the standard attenuation coefficients. However, in realistic applications, the attenuation coefficients may deviate from the standard values. For example, a fatty body may have lower attenuation coefficients than the standard tissue. Thus, it is necessary to investigate the robustness of the proposed method against the assigned attenuation coefficients. In this evaluation, the assigned attenuation coefficients were scaled by 95\% and 105\% intentionally. The results with the mismatched attenuation coefficients were compared to that without variations.

In order to further evaluate the effect of the spectrum model on scatter correction, we have also corrected the thorax phantom using mismatched spectra on purpose. Specifically, two mismatched spectra (1 mm aluminum softer and harder than the true spectrum) were used to test the effect of spectrum model. Results obtained using the mismatched spectra (labeled as A and B) were compared to that without variations by quantitatively measuring the ROIs labeled on the inserts.

In addition, a high frequency bar pattern has been added to the thorax phantom to test the performance of the method. Due to the huge computational cost, we have generated the data in a synthetic fashion. We first forward projected the high-frequency structure (bar pattern) to obtain a set of projection data, which was then added to the previously generated raw total projection data. In this case, we have assumed that the influence of the bar pattern on the scatter distribution is negligible (the bar pattern is small compared to the phantom). The attenuation coefficient of the bar pattern is consistent with the attenuation coefficient of the bone structure in the CT image. To obtain the ground truth, the bar pattern was also added to the primary CT image. The scatter corrections were performed in image domain.

\subsection{Physical phantom experiments}

The proposed algorithm was also evaluated for experimental data of a Catphan600 phantom (The Phantom Laboratory, Salem, NY); scanned using a CBCT on-board imager (Varian 2100EX System, Varian Medical Systems, Palo Alto, CA). The acquisition parameters of the experimental scan are listed in Table~\ref{tab:exparas}. A total of 678 projections were evenly acquired in a 360 degree rotation with $2\times2$ binning and without bow-tie filter. Both, wide collimation and narrow collimation modes were applied with the same scan parameters where the narrow collimation is considered as the scatter-free reference for comparison. During the correction, the 100 kVp polychromatic spectrum was estimated using the raw projection data of a narrow collimated water tank phantom. Quantitative analyses including line profiles as well as the mean values over all pixels in the difference images (primary minus corrected) were performed. 

\begin{table}
\centering
\caption{Acquisition parameters for the experimental phantom scan.}
\label{tab:exparas}
\begin{tabular}{ll} 
\hline
\bfseries  Parameter & \bfseries Value \\
\hline
  Source to detector distance & $1500$~mm \\
  Source to isocenter distance & $1000$ mm   \\
  Number of view angle & $678$  \\
  Tube potential & 100 kVp  \\
  Tube current & 20 mA  \\
  Pulse width & 20 ms \\
  Cone angle for narrow scan & $0.5^{\circ}$  \\
  Cone angle for wide scan & $10^{\circ}$  \\
  Detector size& $397\times298$~mm$^{2}$ \\
  Detector pixel array& $1024\times768$ after $2\times2$ rebinning \\
\hline
\end{tabular}
\vspace{-1em}
\end{table}

\subsection{Patient study}

Patient data from a human pelvis scan was also used to evaluate the proposed method. This data was acquired in half-fan mode using the state-of-the-art CBCT imaging system of the Varian TrueBeam system (Varian Medical Systems, Palo Alto, CA). The tube potential was set to 125 kVp. It has to be noted that the uncorrected images acquired on the scanner have been processed with built-in scatter and bow-tie correction algorithms. To determine the fixed kV spectrum used in the correction, we choose aluminum as the filtration material and tuned the filtration thickness (i.e., extend or reduce the thickness of the aluminum filter), until a reasonable correction was obtained. The spectrum is not estimated in this case with the proposed method because a bow-tie filter is employed in the raw data acquisition. Thus, in order to accurately model the polychromatic projection process, a spectrum along each fan angle would have to be estimated. This would make the whole scatter correction procedure much more complex. In addition, built-in scatter and other correction algorithms impact the spectrum estimation and should be taken into account. However, this is difficult in realistic applications since most of commercial algorithms are proprietary. Hence, we corrected the patient case in image-domain and tuned the spectrum to the optimum. A three-component segmentation (adipose, tissue, and bone) was performed during the correction.  
To evaluate the results, quantitative measurements of contrast and noise in tissue and adipose were performed, where contrast is defined as,
\begin{equation}\label{equ:contrast}
Contrast=\left|\frac{HU_{T}-HU_{B}}{HU_{B}}\right|,
\end{equation}
here $HU_{T}$ and $HU_{B}$ are mean signal intensities of the ROI on target (tissue or adipose) and background (air).


\section{Results} \label{sec:sections}

\subsection{Monte Carlo simulations}

\subsubsection{Calculation in projection domain and image domain}
\begin{figure}[t]
     \centering
     \includegraphics[width=2.5in]{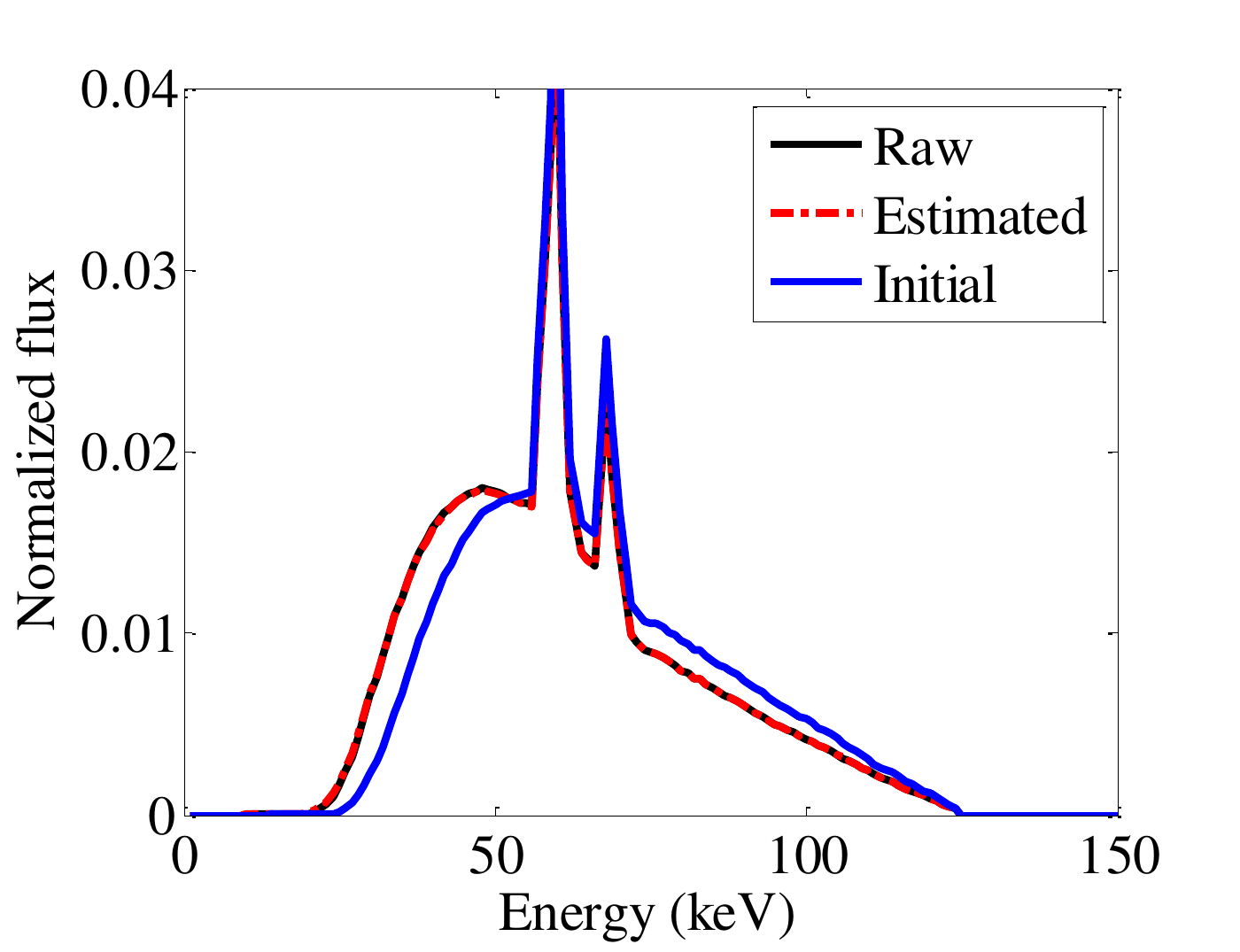}
     \caption{The 125 kV polychromatic spectrum estimated using a water phantom. The initial guess for the spectrum recovery problem corresponds to the hardest model spectrum. }
     \label{fig:spekEstimation}
 \end{figure}

In order to perform the polychromatic reprojection, we first recover the 125 kV spectrum using 10 model spectra which are generated using the Spektr software. Fig.~\ref{fig:spekEstimation} shows the spectrum estimated using the water phantom, together with the true spectrum and the initial guess. Mean energy difference and  normalized root mean square error between the estimated spectrum and the true spectrum is 0.031 keV and $0.42\%$, respectively.

\begin{figure}[t]
     \centering
     \includegraphics[width=3.5in]{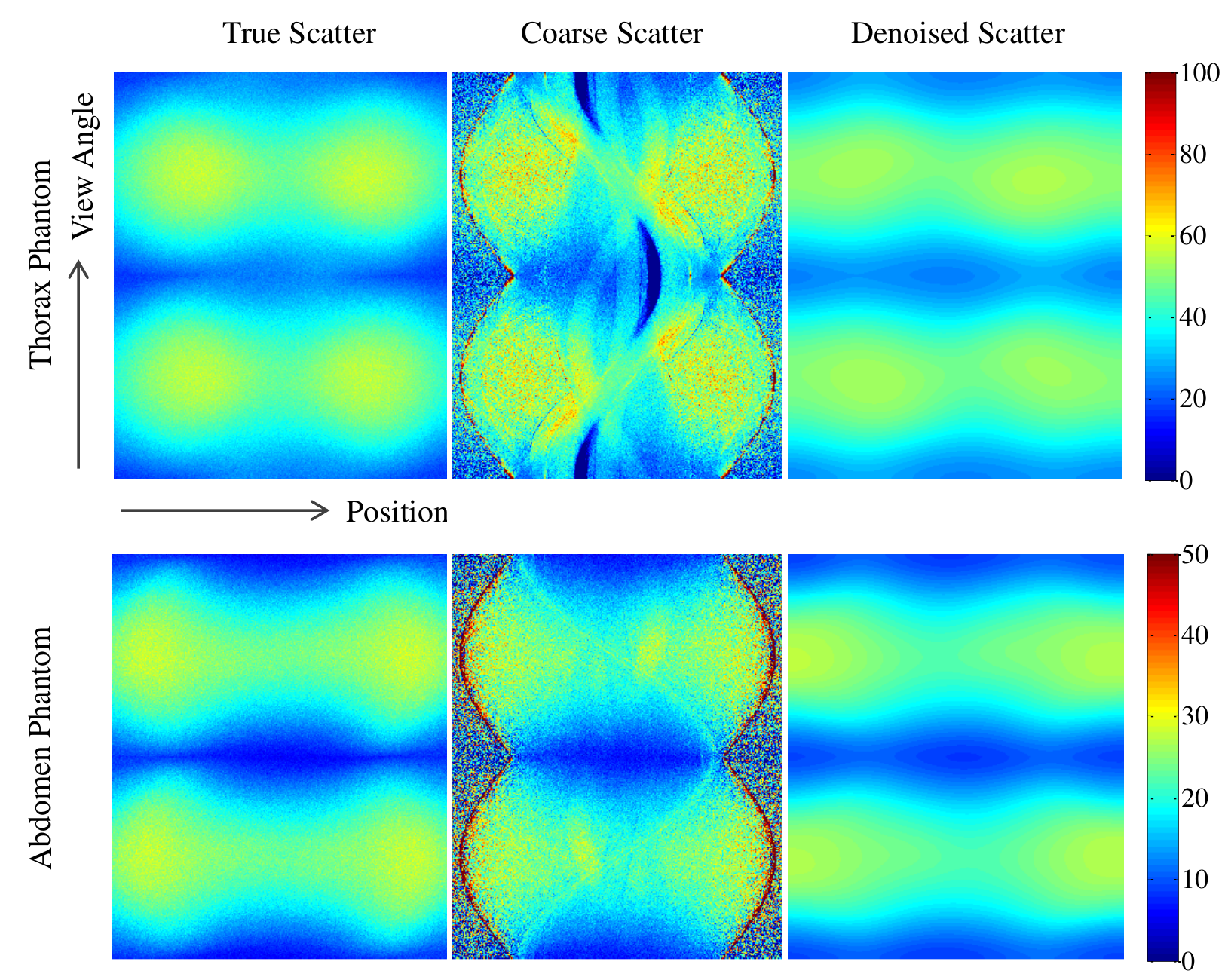}
     \caption{1D projections of true scatter, coarse scatter, and denoised scatter for an anthropomorphic thorax phantom and an abdomen phantom.}
     \label{fig:scatterProfile}
 \end{figure}

Fig.~\ref{fig:scatterProfile} shows the 1D projections of the MC reference scatter, coarse scatter and the denoised scatter for the anthropomorphic thorax phantom and the abdominal phantom. The denoised scatter distributions are obtained with $\beta=100$ and $K=500$. As can be seen, the coarse scatter distributions match globally to the true scatter profiles but they contain edge errors and low contrast errors caused by the inaccurately segmented structure contours. In the denoised scatter, these errors are smeared out and fit the true reference scatter quite well. Note that, for the same incident X-ray intensity, the magnitude of the scatter signal of the thorax phantom is much larger than the magnitude of the scatter signal of the abdomen phantom, as the abdomen phantom has a larger attenuation  and can absorb more scatter and primary photons.

 \begin{figure}[t]
     \centering
     \includegraphics[width=3.5in]{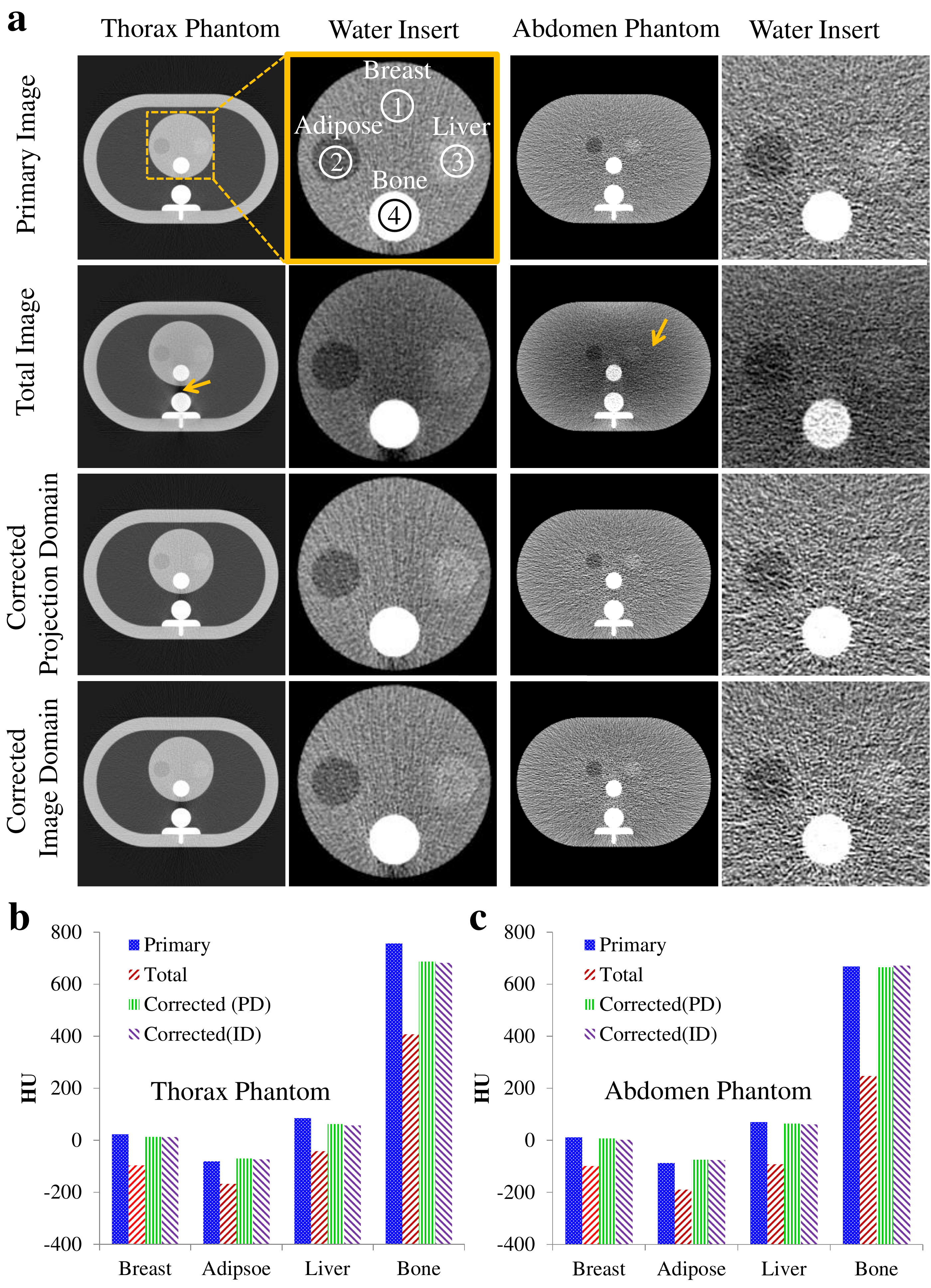}
     \caption{Results of scatter corrections for the MC simulation data of the thorax phantom and the abdomen phantom in both projection domain (PD) and image domain (ID). (a) CT images. Display window: [-1200 HU, 500 HU] for thorax phantom images and [-300 HU, 300 HU] for the water insert images and the abdomen phantom. (b-c) Results of HU numbers at labeled ROIs of the water insert of  the thorax phantom and the abdomen phantom before and after scatter correction.}
     \label{fig:domains}
 \end{figure}

Fig.~\ref{fig:domains}(a) shows the results of the scatter correction using the proposed method for the MC simulation data of the thorax phantom and the abdomen phantom within both projection domain and image domain. The primary images were reconstructed using primary projections and they are therefore scatter-free images (1st row of Fig.~\ref{fig:domains}(a)). The total images (2nd row of Fig.~\ref{fig:domains}(a)) were reconstructed using total projections (primary signal plus scatter signal). Scatter induced shading artifacts and streaks are clearly visible in the images. Scatter corrections were performed in both projection domain (3rd row of Fig.~\ref{fig:domains}(a)) and image domain (4th row of Fig.~\ref{fig:domains}(a)); showing that shading artifacts were significantly reduced in both cases. Fig.~\ref{fig:domains}(b) and (c) depict the HUs of breast, adipose, liver, and bone inserts (shown in Fig.~\ref{fig:domains}(a)) of the primary, total, and scatter corrected images. Compared to the scatter-free primary images, the HUs of the total images were greatly reduced for both the thorax phantom and the abdomen phantom. After scatter correction in either projection domain or image domain, the HUs were successfully recovered.

Noise levels of the MC simulation studies of the thorax phantom and the abdomen phantom are depicted in Table~\ref{tab:noiseLevel}. As can be seen, the total images have the lowest noise levels as they have the most photon counts. After scatter correction, noise levels increase because the subtracted denoised coarse scatter is a low frequency signal and noise is left in the corrected projection data.

\begin{table}
\vspace{0em}
\caption{Noise levels of the MC simulation studies of the thorax phantom and the abdomen phantom.}
\label{tab:noiseLevel}
\begin{center}
\begin{tabular}{p{0.08\textwidth}<{\centering}p{0.08\textwidth}<{\centering}p{0.08\textwidth}<{\centering}p{0.08\textwidth}<{\centering}p{0.10\textwidth}<{\centering}p{0.10\textwidth}<{\centering}} 
\toprule
 Phantom &  ROIs & Primary & Total & Corrected (PD) & Corrected (ID) \\
\hline
\multicolumn{1}{ c }{\multirow{3}{*}{Thorax } } & Breast & 43 & 36 & 50 & 51 \\& Adipose & 42 & 36 & 50 & 49 \\
& Liver & 46 & 39 & 52 & 54 \\
\hline
\multicolumn{1}{ c }{\multirow{3}{*}{Abdomen} } & Breast  & 88 & 72 & 111 & 115\\
&Adipose & 93 & 71 & 121 & 126 \\
&Liver & 106 & 73 & 125 & 132 \\
\bottomrule
\end{tabular}
\end{center}
\end{table}

\subsubsection{Segmentation}

\begin{figure}[t]
     \centering
     \includegraphics[width=3.5in]{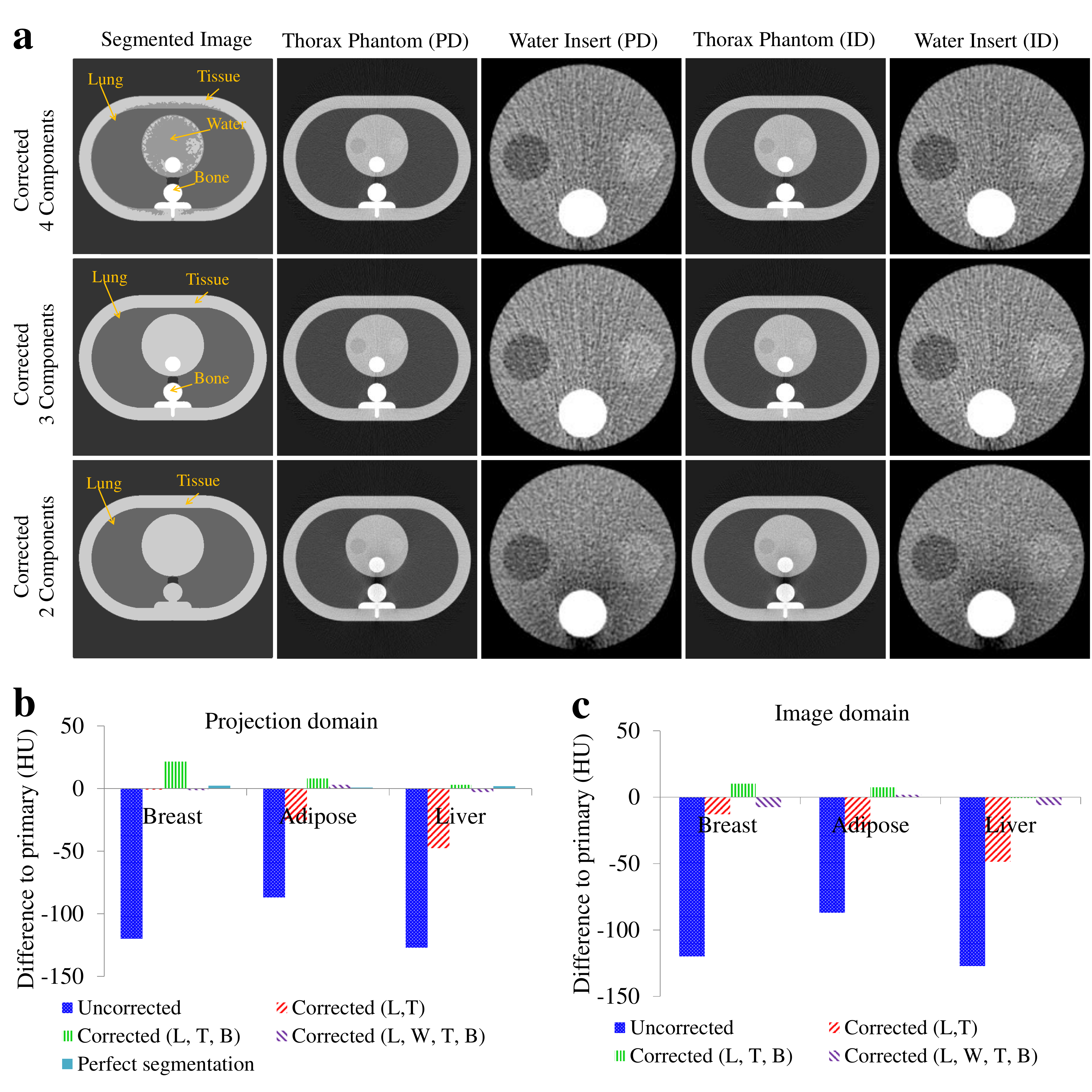}
     \vspace{-1em}
     \caption{The influence of the segmentation methods on the accuracy of the scatter correction is shown for the thorax phantom in both projection domain (PD) and image domain (ID). (a) Results of different segmentation methods. Display window: [-1200 HU, 500 HU] for the thorax phantom images and [-300 HU, 300 HU] for the water insert images. (b-c) Results of the difference to the primary references of the reconstruction values in HU numbers for different ROIs of the water insert.  L, W, T, and B stand for lung, water, tissue, and bone, respectively.}
     \label{fig:segmentaion}
 \end{figure}

The influence of different segmentation methods on the accuracy of the scatter correction for the thorax phantom is depicted in Fig.~\ref{fig:segmentaion}. The scatter correction was performed in both projection domain and image domain. For the four components segmentation, where lung, water, tissue, and bone were identified with thresholds -802 HU, -408 HU, -40 HU and 242 HU, the adipose low contrast insert was missed due to the presence of a high MC noise level and shading artifacts. Besides, a part of the lung area was also identified as air because of the presence of streak artifacts which is shown between the bone insert and the spine. However, in this case, scatter artifacts were still significantly reduced (Fig.~\ref{fig:segmentaion}(a), 1st row). For the three components segmentation case, where lung, tissue, and bone were identified with thresholds -855 HU, -382 HU and 229HU, the corrected images are still acceptable (Fig.~\ref{fig:segmentaion}(a), 2nd row). There are residual shading artifacts for the corrected images when only two components segmentation was performed, where only lung and tissue were identified (Fig.~\ref{fig:segmentaion}(a), 3rd row).

Fig.~\ref{fig:segmentaion}(b-c) show the HU difference between the corrected images for the different segmentation methods and the primary image for the projection and image domain implementation, respectively. The uncorrected images were reconstructed using total projections. Corrected (L, T) (corresponds to the 3rd row of Fig.~\ref{fig:segmentaion}(a)) means the proposed scatter correction method was performed using two components segmentation where lung and tissue were identified, and so on. Compared to the HU differences of the uncorrected image, the differences of the corrected images are significantly reduced. Note that when the two components segmentation was performed, the spine and bone insert were identified as tissue. Thus, the photon counts of the polychromatic reprojection data were overestimated, underestimating the scatter contribution in both coarse and denoised scatter. In this case, there are residual scatter artifacts (Fig.~\ref{fig:segmentaion}(a), 3rd row) and the HUs of the corrected images are less than that of the primary images. When the three components segmentation was performed, the water insert was identified as tissue, which has a slightly larger attenuation coefficient. Thus the photon counts of the polychromatic reprojection were underestimated, overestimating the scatter contribution in both coarse and denoised scatter. In this case, the corrected images are slightly overcorrected and the HUs of the corrected images are higher than the HUs of the primary images. The perfect segmentation result shows that minor segmentation errors in the four component segmentation have no significant influence in the correction.

\begin{figure*}[t]
     \centering
     \includegraphics[width=\textwidth]{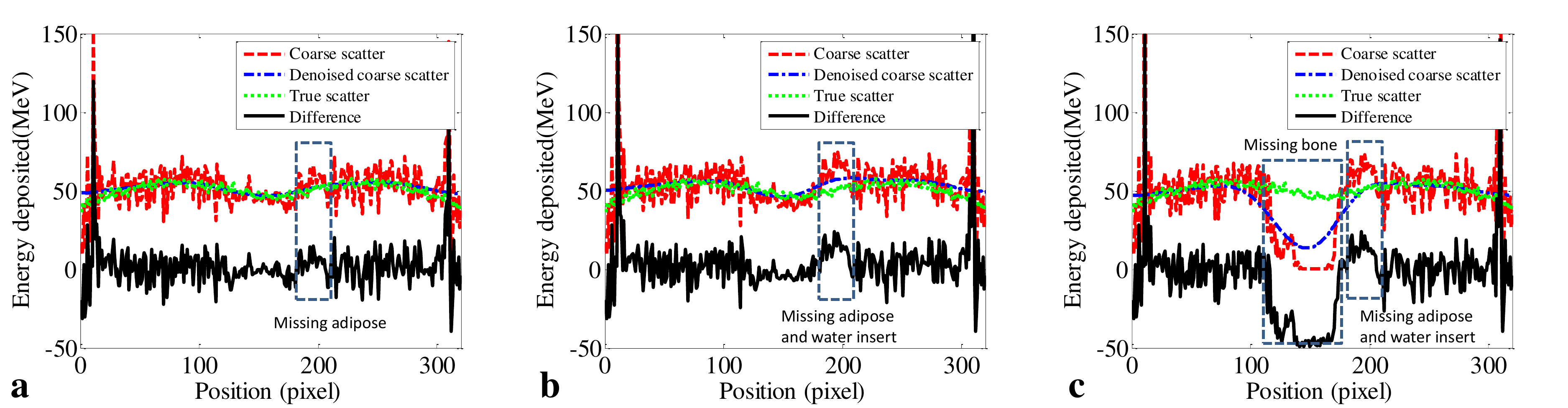}
     \vspace{-1em}
     \caption{Line profiles of the true scatter, coarse scatter, denoised coarse scatter and the difference between the coarse scatter and true scatter for the thorax phantom with (a) four, (b) three, and (c) two components segmentation methods.}
     \label{fig:segmentaionLineProfile}
 \end{figure*}

To further demonstrate that the proposed scatter correction method is not very sensitive on segmentation errors, line profiles of the true, coarse, and denoised scatter as well as the difference between the coarse and true scatter of the thorax phantom using four-, three- and two-components segmentation, are depicted in Fig.~\ref{fig:segmentaionLineProfile}. When four components segmentation was applied (shown in Fig.~\ref{fig:segmentaionLineProfile}(a)), the denoised scatter profile fits the true scatter quite well, although the adipose was identified as water. This can be attributed to the mitigating effect of the denoising procedure on the error (dashed box). The incorrect segmentation caused error was reduced after denoising. The same effect can be seen in Fig.~\ref{fig:segmentaionLineProfile}(b) where the three components segmentation was applied. In this case, both adipose and water inserts were identified as tissue, enhancing the error. However, the denoising procedure significantly reduced the errors. This is the reason why the three components segmentation still yields acceptable results (Fig.~\ref{fig:segmentaion}(a), 2nd row). For the two components segmentation, where bone, adipose and water inserts were all identified as tissue (Fig.~\ref{fig:segmentaionLineProfile}(c)), the denoising procedure partly compensated the error. However, the coarse scatter was still significantly underestimated, causing residual scatter artifacts (Fig.~\ref{fig:segmentaion}(a), 3rd row).

Based on these results, the four components segmentation is used for all further simulation studies.

\subsubsection{Parameters selection}

 \begin{figure}
     \centering
     \includegraphics[width=3.5in]{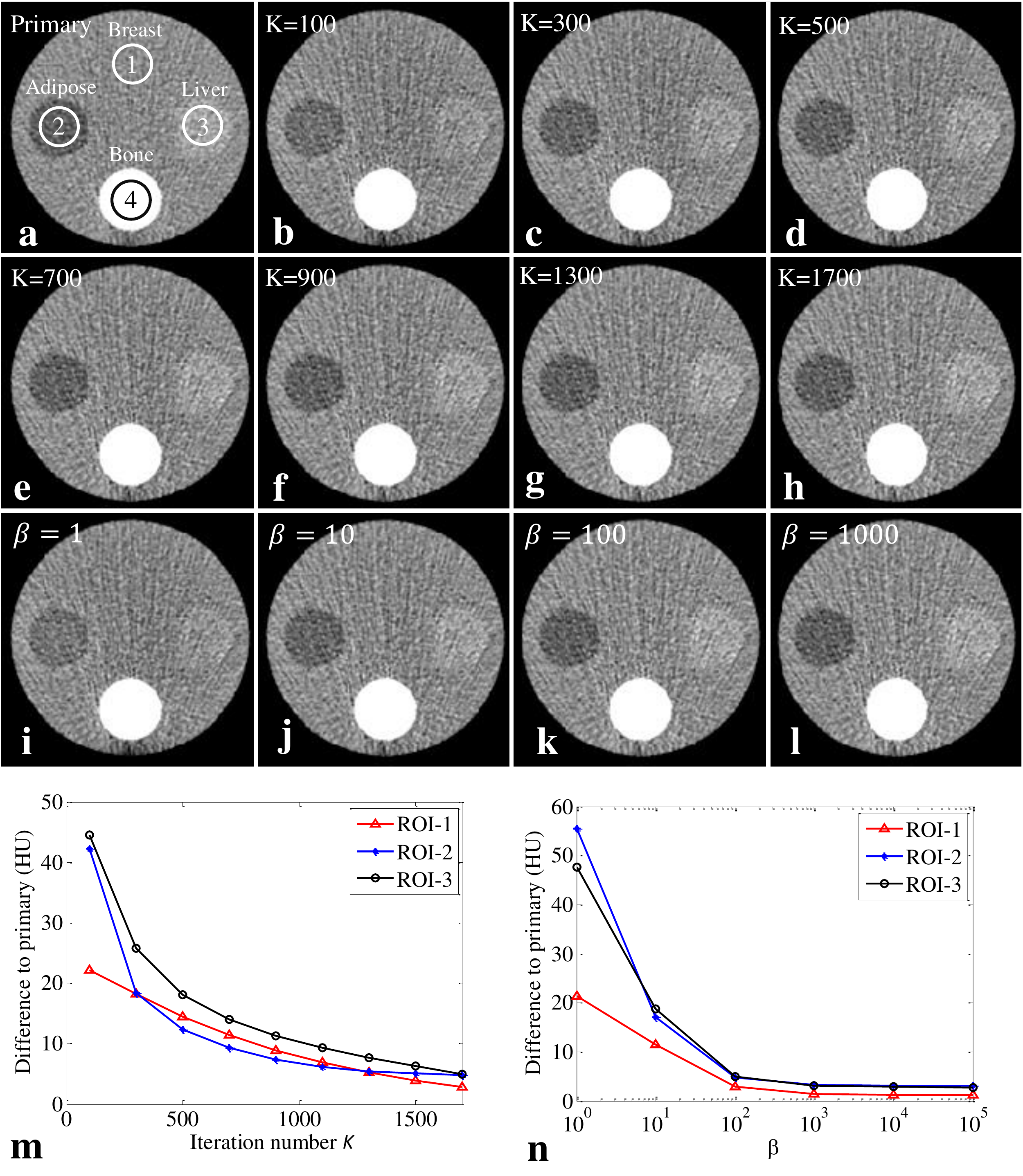}
     \vspace{-1em}
     \caption{The influence of the iteration number $K$ and $\beta$ on the accuracy of the scatter correction is shown for the water insert of the thorax phantom. (a) Image reconstructed using primary signal. (b-h) Scatter correction results using different number of iteration $K$. (i-l) Scatter correction results using different $\beta$. (m-n) Difference to the primary references of the HU numbers of the low contrast inserts after scatter correction using different $K$ and $\beta$. Display window for the images: [-300 HU, 300 HU].}
     \label{fig:iterationNum}
 \end{figure}

In Fig.~\ref{fig:iterationNum}, the influence of the iteration number $K$ and parameter $\beta$ on the accuracy of the scatter correction is shown for the water insert of the thorax phantom. As can be seen, the HUs of the low contrast inserts were gradually recovered as $K$ was increased from 100 to 1700, while keeping $\beta$ as 10000 (Fig.~\ref{fig:iterationNum}(b-h)). This is because the high frequency content in the coarse scatter is smeared out as $K$ is increased, leaving only low frequency content in the profiles which should represent the true scatter profile. Thus, the proposed method can regain HU accuracy even though the segmentation was badly performed. The HU differences between the reference primary image and the corrected image of the low contrast inserts are shown in Fig.~\ref{fig:iterationNum}(m). The differences to the primary values are significantly reduced as $K$ increases. When $K=1700$, the HU differences are less than 5 HUs for all of the three low contrast inserts.

Similar results can be seen for the parameter $\beta$ which determines the relative importance of the data fidelity term and the regularization term in (\ref{equ:energyfunction}), i.e. the low contrast visibility were gradually recovered as $\beta$ is increased (Fig.~\ref{fig:iterationNum}(i-l)) while keeping the iteration number $K$ as 1700. The differences to the primary HUs are greatly reduced as $\beta$ increases (Fig.~\ref{fig:iterationNum}(n)), i.e. the regularization term is more pronounced. When $\beta$ is larger than 1000, the HU differences to the primary HUs are negligible, suggesting the error content of the coarse scatter is well smeared out and the proposed method yields a low contrast resolution that can be compared to the primary images.

\subsubsection{Robustness evaluation}

\begin{figure}
     \centering
     \includegraphics[width=3.5in]{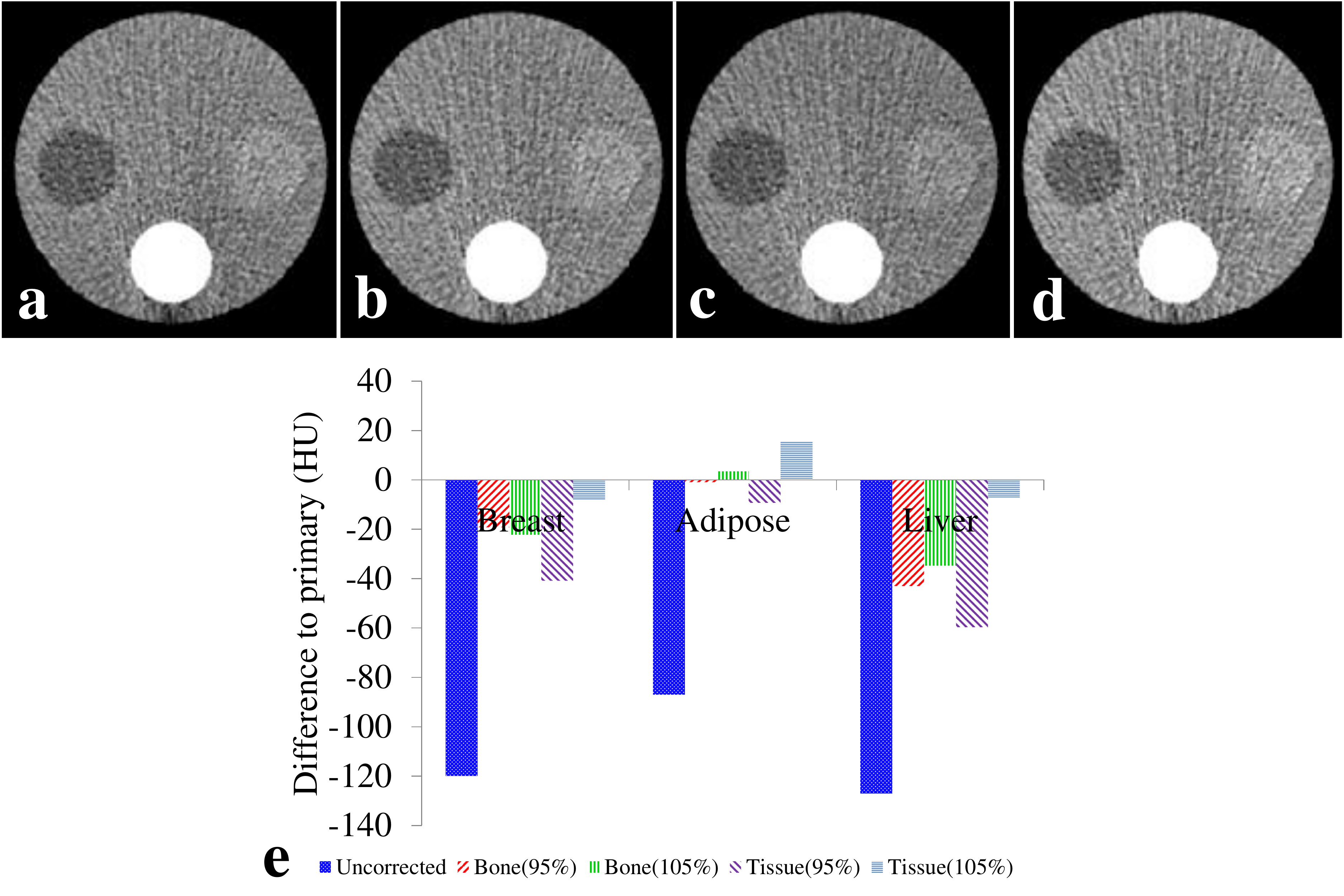}
     \vspace{-1em}
     \caption{Results of the water insert of the thorax phantom using mismatched attenuation coefficients. (a) and (b) Correction with 95\% and 105\% of the standard bone attenuation coefficient, respectively. (c) and (d) Correction with 95\% and 105\% of the standard tissue attenuation coefficient, respectively. (e) Results of the difference to the primary references. Display window for the images: [-300 HU, 300 HU].}
     \label{fig:mismatchedAtten}
 \end{figure}

In this section, we investigate the influence of the mismatched attenuation coefficients on the scatter correction results. Fig.~\ref{fig:mismatchedAtten} shows the results of the water insert of the thorax phantom using mismatched attenuation coefficients. To quantitatively depict the accuracy of HUs after correction with mismatched attenuation coefficients, the difference between the corrected images and the primary reference in HU for the different ROIs of the water insert are calculated (shown in Fig.~\ref{fig:mismatchedAtten}(e)). As can be seen, when the assigned attenuation coefficients of bone were scaled by 95\% and 105\%, a significant removal of scatter artifacts is still achieved; reducing the HU error from -120 HUs to below -40 HUs. When the attenuation coefficient of tissue was scaled by 105\%, the HU error can be reduced lower than 20 HU. However, the 95\% scaled tissue still misses the correct HU in the liver by -60 HU.


Table~\ref{tab:robustTest} shows quantitative measurements of the ROIs labeled on breast, adipose, and liver inserts for scatter corrections using mismatched source spectra. Compared to the HU values of the total image, all of the three corrections significantly recover the HU accuracy. However, corrections with the mismatched spectra show degraded quantitative accuracy, as expected. When the softer spectrum A is used, the method tends to underestimated the coarse scatter, yielding an under-correction. On the contrary, the method tends to over-correct the image when the harder spectrum B is used.

\begin{table}
\vspace{0em}
\caption{Results of scatter correction using mismatched spectra A and B. HU values of the ROIs that labeled on the inserts are measured.}
\label{tab:robustTest}
\begin{center}
\begin{tabular}{cp{0.10\textwidth}<{\centering}p{0.10\textwidth}<{\centering}p{0.10\textwidth}<{\centering}p{0.10\textwidth}<{\centering}p{0.10\textwidth}<{\centering}} 
\toprule
ROIs &Primary & Total & Standard Correction & Correction with A & Correction with B\\
\hline
\rule[-1ex]{0pt}{3.5ex}  Breast & 23 & -97 & 22 & 14 & 41 \\
\rule[-1ex]{0pt}{3.5ex}  Adipose & -81 & -168 & -78 & -82 & -72  \\
\rule[-1ex]{0pt}{3.5ex}  Liver & 85 & -42 & 82 & 73 & 94  \\
\bottomrule
\end{tabular}
\end{center}
\end{table}

 \begin{figure}
     \centering
     \includegraphics[width=3.0in]{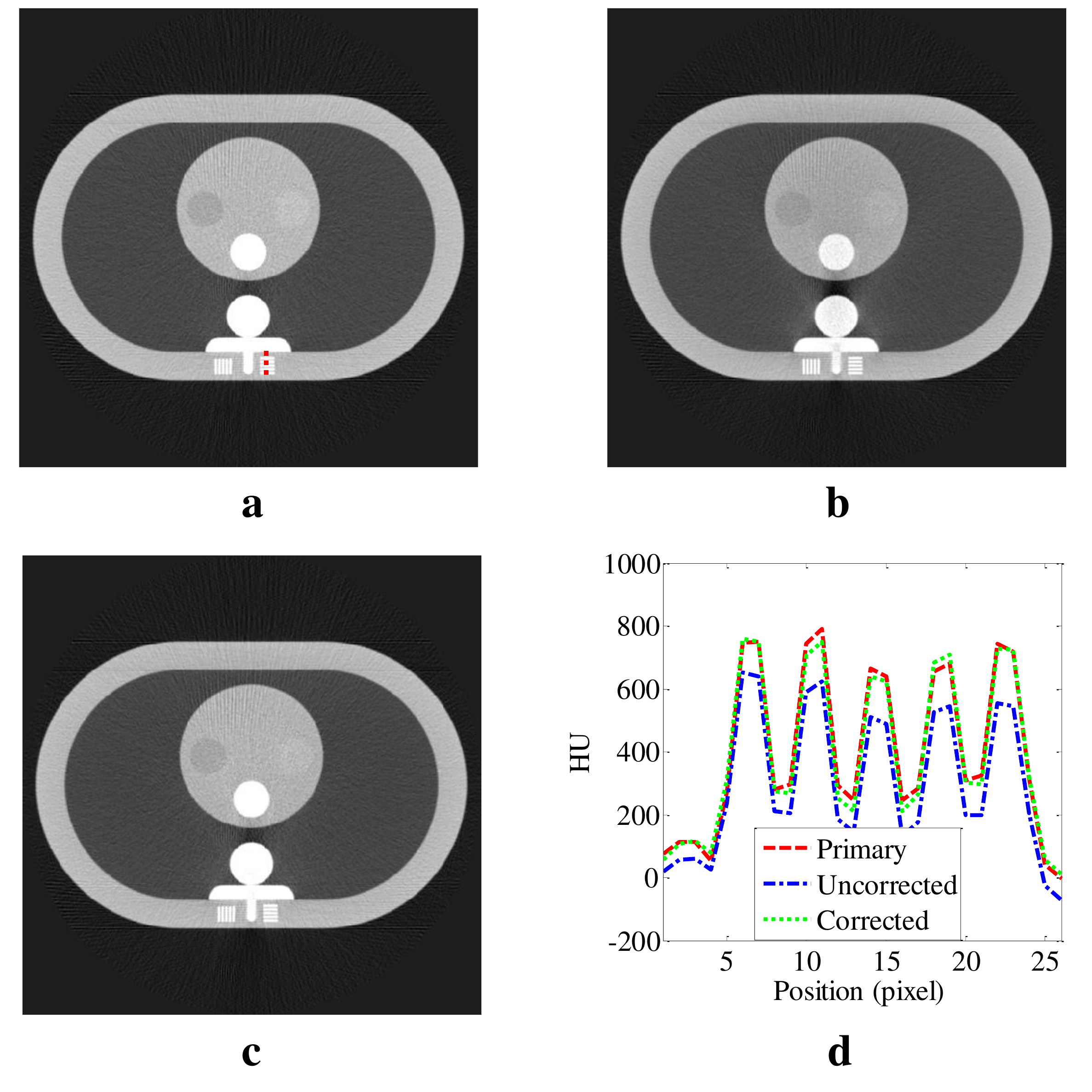}
     \vspace{-1em}
     \caption{Results of the the thorax phantom with bar pattern. (a) Primary image, (b) total image, (c) corrected image, (d) line profiles (the red line in (a))of images. HU accuracy is greatly improved and spatial resolution is well preserved after scatter correction. Display window for the CT images: [-1200 HU, 500 HU].}
     \label{fig:barpattern}
 \end{figure}

Fig.~\ref{fig:barpattern} shows results of the scatter correction for the thorax phantom with bar pattern. It is visible that shading artifacts were significantly reduced. Line profile suggests that HU accuracy is greatly improved and spatial resolution is well preserved after correction.

\subsection{Experimental phantom results}
 \begin{figure}[t]
     \centering
     \includegraphics[width=3.5in]{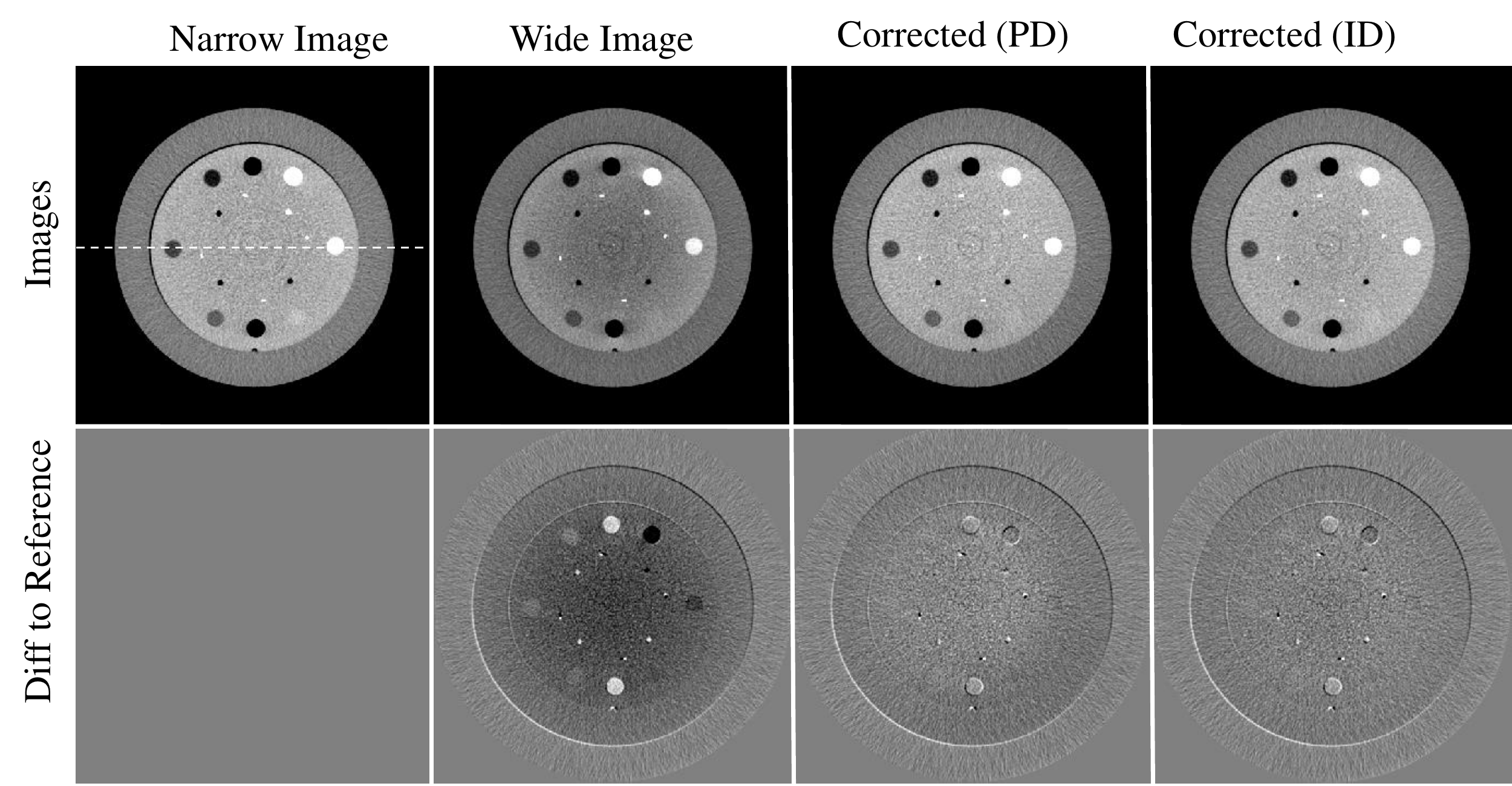}
     \caption{Catphan$^{\textregistered}$600 phantom with and without scatter correction. The difference images show each image subtracted with the narrow collimation image. Display window: [-200 HU, 200 HU] for both the CBCT images and the difference images.}
     \label{fig:catphan}
 \end{figure}

 \begin{figure}[t]
     \centering
     \vspace{-1em}
     \includegraphics[width=3.0in]{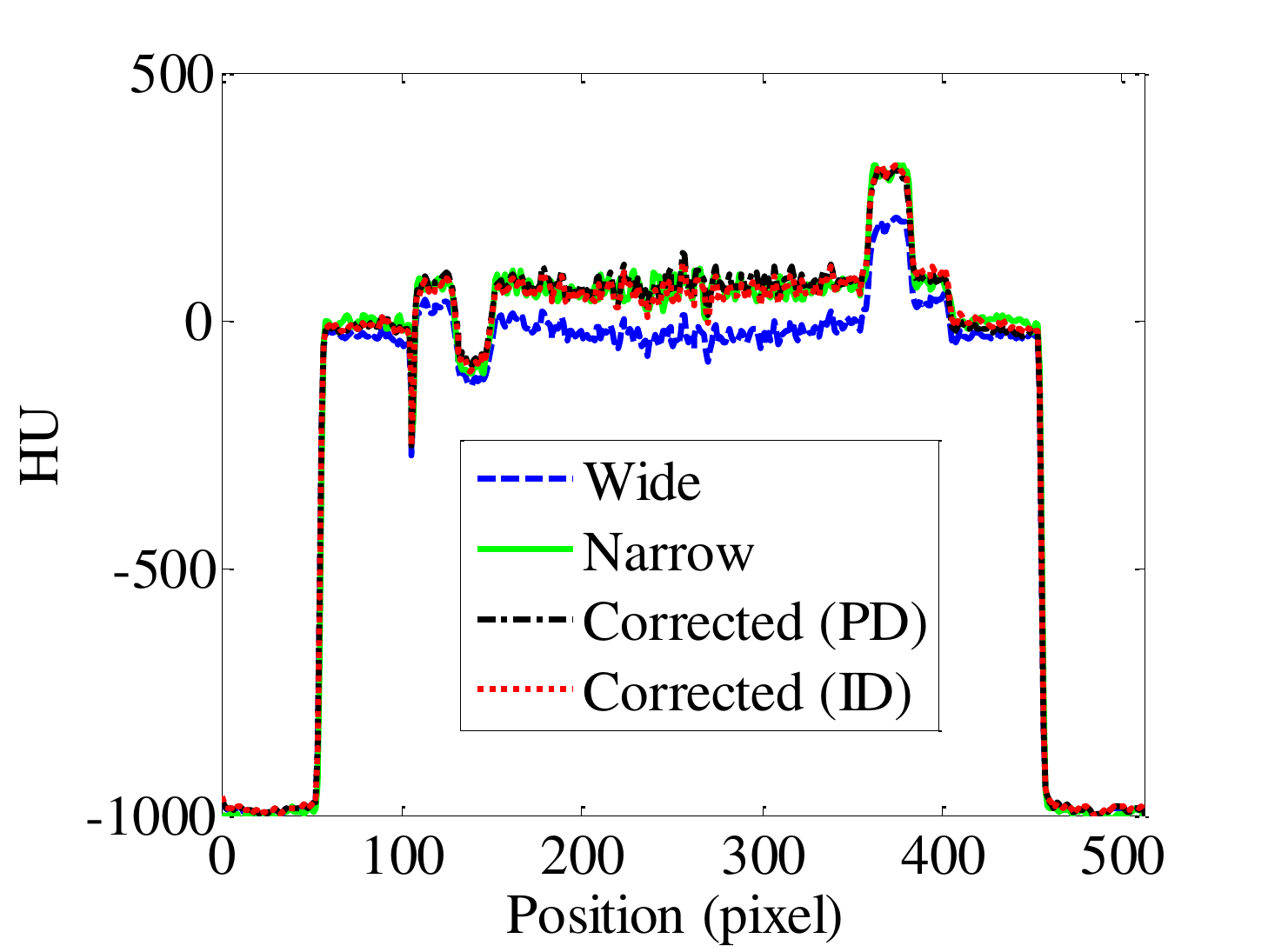}
     \vspace{-0.5em}
     \caption{Line profiles of the Catphan$^{\textregistered}$600 phantom without and with scatter correction using both projection domain (PD) and image domain (ID) methods.}
     \label{fig:catphanLine}
 \end{figure}

Fig.~\ref{fig:catphan} shows the CT images and the corresponding difference images of the Catphan$^{\textregistered}$600 phantom with and without scatter correction. The narrow image was reconstructed using narrow collimation projection data which was considered as the scatter-free data, thus the narrow image was served as reference for comparison. The wide image was reconstructed using wide collimation projection data and it contains scatter radiation. In the reconstructed image, shading artifacts are visible. The difference image depicts that the HU accuracy is reduced by the presence of artifacts. Shading artifacts were greatly reduced in the scatter corrected images for both the projection and image domain implementations. Note that the wide collimation scan and the narrow collimation scan are two independent scans and the registration can not be perfect, thus there are edge fringes in the difference images. Line profiles (illustrated as dashed line in Fig.~\ref{fig:catphan}) of the Catphan$^{\textregistered}$600 phantom with and without scatter correction are shown in Fig.~\ref{fig:catphanLine}. As can be seen, the HUs of the wide collimation image were significantly reduced by scatter radiation. After correction with the proposed method, scatter induced HU reduction was successfully recovered and the profiles match the reference profile quite well. The mean value over all pixels in the difference image is reduced from
-21.8 HU to -0.2 HU and 0.7 HU for PD and ID correction, respectively.

\subsection{Patient study}

 \begin{figure}[t]
     \centering
     \includegraphics[width=3.5in]{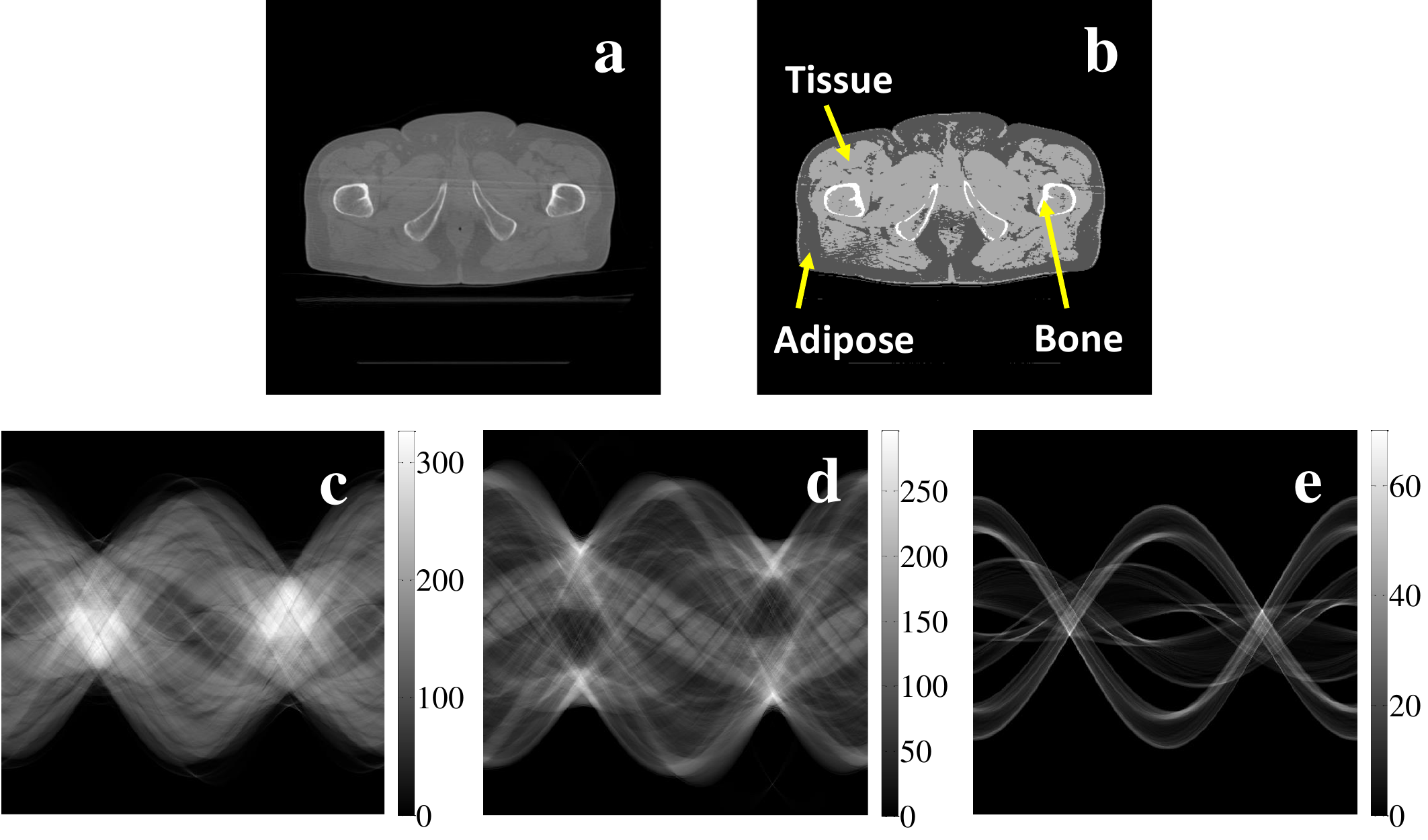}
     \caption{Uncorrected image, segmented image and corresponding propagation path length in the segmented materials for a pelvis scan. (a) Uncorrected pelvis image. (b) Three components (adipose, tissue and bone) segmentation image. Propagation path length in (c) tissue, (d) adipose, and (e) bone. Note that the unit for propagation path length is millimeter.}
     \label{fig:reproj}
 \end{figure}

 \begin{figure}[t]
     \centering
     \includegraphics[width=3.5in]{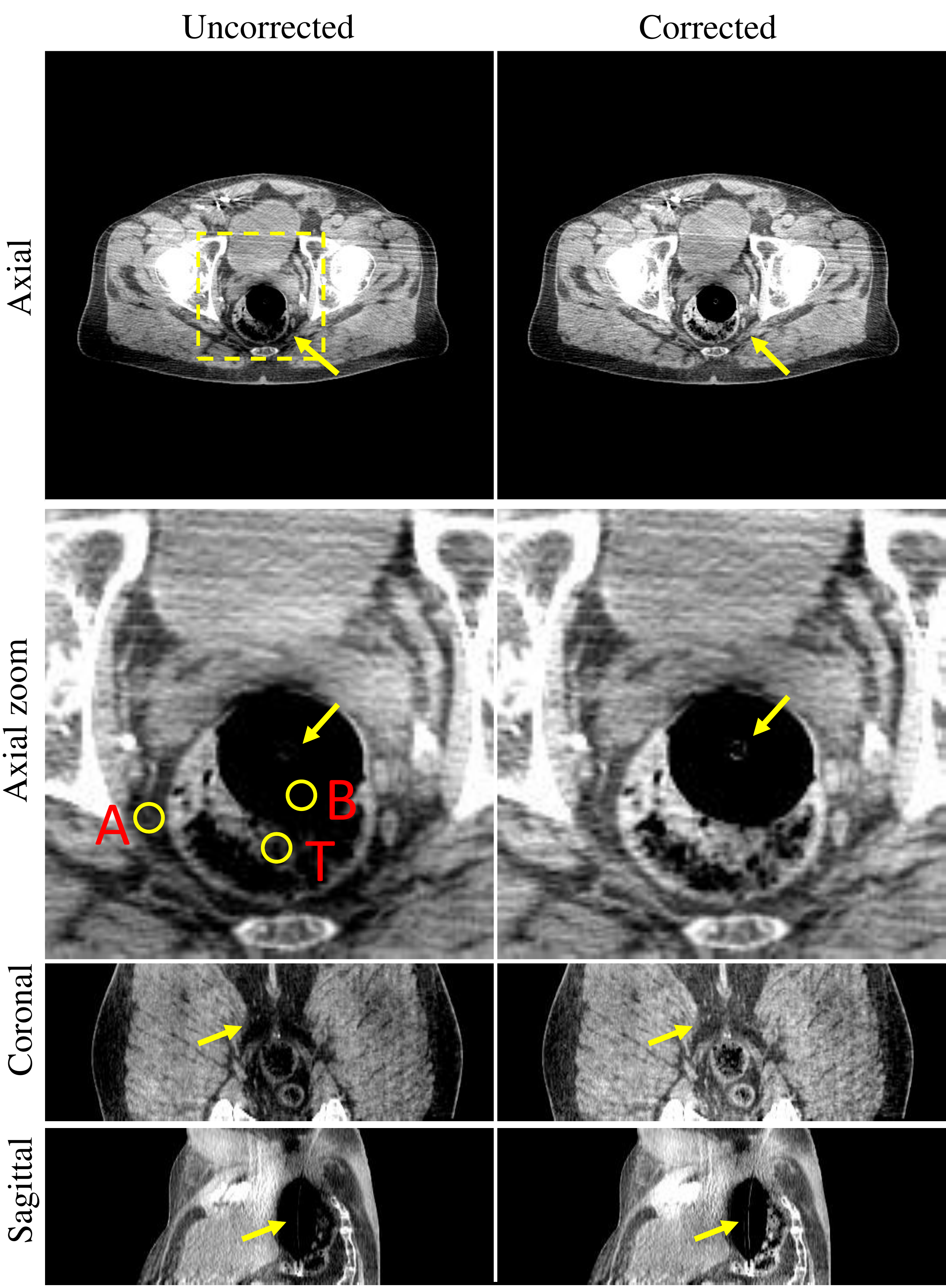}
     \caption{Scatter correction for a pelvis CBCT scan using the kV imaging system of the Varian TrueBeam system. Axial view, coronal view and sagittal view are depicted respectively. The three ROIs (labeled as A-adipose, B-background, and T-tissue) are used to calculate contrast before and after scatter correction. Display window: [-200 HU, 100 HU] for all images.}
     \label{fig:clinicalData}
 \end{figure}
The uncorrected image, segmented image and the PPLs for the segmented components for the pelvis scan are depicted in Fig.~\ref{fig:reproj}. Fig.~\ref{fig:clinicalData} shows CT images of an \emph{in vivo} pelvis scan with the kV CBCT imaging system of the Varian TrueBeam system (Varian Medical Systems, Palo Alto, CA). It can be seen that a dark region or "black hole" is present in the uncorrected images. In order to obtain a reasonable correction, it is expected that adipose and tissue have uniform HUs in the dashed box in the first row of Fig.~\ref{fig:clinicalData}. However, visual inspection of the corrected images show nonuniformity and residual scatter artifacts in the form of shading areas. Hence, the spectrum was gradually tuned until adipose and tissue uniformity was achieved. After scatter correction, the HUs of the dark region and the missing anatomical structures are successfully recovered. Both tissue and adipose contrasts have been improved from 0.85 and 0.85 to 0.96 and 0.90, respectively. Note that due to the limitation of access to the raw projection data, the correction of this clinical case was performed in image domain and the uncorrected images have been preprocessed using the built-in bow-tie and scatter correction algorithms implemented by the vendor.

\section{Discussion}
\label{sec:discussion}
The proposed method can be used either prospectively or retrospectively for improved CBCT imaging. The computational demand of the method depends mainly on the polychromatic reprojection and the denoising procedure. The computing time of the former case is similar to that of the backprojection.  The denoising procedure can be implemented in a parallel fashion using GPU acceleration~\cite{pratx2011,jia2012}. It was found that it took about 1~min to correct a typical Varian clinical dataset ($512\times512\times81$) using a NVIDIA GeForce GTX~480 card, thus the method is suitable for clinical applications. It is worthwhile to mention that this method can be applied to both flat detector-based CBCT and spiral CBCT scanners (without ASG), especially to dual source dual energy CT scanners where convolution-based techniques do not work for cross-scatter~\cite{petersilka2010}. The method, which allows to correct the scatter artifacts in scale of minutes, could potentially be used in spiral CBCT (without anti-scatter grid) and flat detector CBCT.


It has to be noted that the selection of the denoising parameters depends on the segmented image. In this study, since the noise level of the MC simulation data is very high, relatively large $K$ and $\beta$ values were used to smooth the segmentation error and to yield a denoised scatter distribution that fits to the true scatter well enough. To our belief, most of the clinical images have a much lower noise level and the images are superior for segmentation than the images from the MC simulation data in this study. For example, compared to the MC simulation studies, it only took 280 iterations to yield acceptable results for the \emph{in vivo} data. 

Different from the MC and the experimental phantom studies where dedicated spectrum estimation was performed as a general procedure of system calibration of scatter correction, for the retrospective patient study, no spectral estimation was done to reflect the realistic situation. In the later case, only different filtrations were tested for optimal scatter correction as the details of filtration during the patient scanning was not known. For prospective studies or clinical applications in the future where dedicated scatter free data is not available, spectrum estimation method that is a part of the scatter correction framework may not provide sufficient accuracy. In these cases, other spectrum estimation methods like e.g.~\cite{duisterwinkel2015} can be employed or the spectrum can be estimated from known filtration obtained from the scanning protocol.

Since the subtracted denoised coarse scatter is a low frequency signal, noise is left in the corrected projection data, enhancing the noise level of the CT images. However, the enhanced noise can be efficiently reduced by many existing algorithms, such as the penalized weighted least-squares (PWLS) algorithm~\cite{wang2006penalized,zhu2009noise}. In this study, we have employed the Poisson statistic-based denoising model to refine the coarse scatter signal. Other methods can be also used to refine the coarse scatter. For example, a convolution-based scatter model~\cite{zhao2015scatter} and a similar method~\cite{wu2015iterative} that uses the Savitzky-Golay filter to smooth the residual image between the uncorrected image and a template image were proposed recently.

The evaluation studies did not take into account the potential impact of bow-tie filters. When a bow-tie filter is used in data acquisition, the bow-tie specifications need to be incorporated into the reprojection procedure for the projection domain implementation. In principle, the spectrum estimation method~\cite{zhao2015} can calculate spectra along different fan angles, thus the scatter correction method can be expanded to work also with bow-tie filters. However, in this case, the polychromatic reprojection procedure would be much more complicated as each fan angle corresponds to a different spectrum and one may want to use a single effective spectrum. For the patient study, the scatter correction was performed in image-domain and the spectrum was tuned to yield reasonable results. Here, the spectrum can be regarded as an effective spectrum that incorporates all of the effects of data acquisition, and built-in correction algorithms.

Comparing the HUs of the corrected images and the primary images in Fig.~\ref{fig:domains}(b-c), and the HUs of the corrected images and the narrow image in Fig.~\ref{fig:catphanLine}, one can see that the proposed method yields quantitative values close to the reference ones. The interpretation of the narrow image as a scatter-free reference image might be questionable as a certain amount of scatter remains. However, the scatter amplitude is negligible compared to that in the wide scan. Nevertheless, the HU consistency demonstrates the potential of the method to provide quantitative CBCT imaging with flat detectors.

In image-guided radiation therapy (IGRT), diagnostic multidetector CT (MDCT) images are acquired for treatment planning prior to the start of patient treatments. Hence, a MDCT-based shading correction method was proposed for flat detector CBCT systems~\cite{niu2010,niu2012}. Clinical cases processed using this method have shown promising results. The proposed method has a workflow similar to the MDCT-based method. Compared to the MDCT-based method where the primary projections are generated by a forward projection of the registered MDCT images, the proposed method generates the primary projections using a polychromatic reprojection with the segmented uncorrected image, suggesting no need of an additional MDCT acquisition beforehand. 

One of the potential limitations of the proposed method is that it assumes a linear signal response for detector pixels and does not consider dynamic range limitation. In practice, the dynamic range of the flat detector is limited, and thus the detector pixels may work in a nonlinear response region, especially for pixels exhibiting photon starvation or pixels with saturated X-ray flux. As a result, the estimated primary signal $\hat{I}_{p}$ cannot match the real primary data, which may cause negative values in the scatter estimate. Thus a non-negative constraint is usually applied and the scatter fraction clipping technique~\cite{Sun2010} is also employed to partially compensate the limitation.

\section{Conclusions}
In summary, this work investigates a novel scatter correction method for high quality CBCT imaging. In this technique, after the initial CBCT image reconstruction with the raw projection data, the uncorrected images were segmented for the purpose of subsequent polychromatic reprojection and scatter estimation. The scatter correction then proceeds in two steps: (1) estimating the coarse scatter profile by computing the difference between the measured raw data and a polychromatic reprojection of a segmented image volume, where the energy spectrum for the polychromatic reprojection calculation was obtained by an indirect transmission measurement-based spectrum estimation method~\cite{zhao2015}; and (2) improving the accuracy of the scatter radiation distribution by applying a denoising algorithm. A detailed evaluation study indicated that the scatter artifacts, such as cupping and streaks, were mitigated significantly after correction with the proposed method. The results also demonstrated that a significant increase in image uniformity and HU accuracy were achieved after correction. On the practical aspect, the proposed method requires minimal increase in computational cost with no modification in system hardware or clinical workflow. When implemented practically, this should lead to a significant clinical impact in image-guided interventions and adaptive radiation therapy treatment planning based on CBCT.


\section*{Acknowledgment}
The authors would like to thank Dr.~Stephen~Brunner and Dr. Jennifer Smilowitz for providing the experimental phantom data and patient data. The authors are grateful to Dr. Kai Niu for his help on GPU implementation. The authors are also grateful to Drs. Orhan Unal, Xuexiang Cao, Guanghong Chen, Jie Tang and Ke Li for their technique support and useful discussions. Lei Xing is supported partially by a grant from NIH/NIBIB (1R01-EB016777).


%
%

%



\end{document}